\documentclass[useAMS,usenatbib]{mnras}
\usepackage{graphicx}
\usepackage{amsmath}
\usepackage{aas_macros}
\usepackage{subfig}
\usepackage{graphics}
\usepackage{times}
\usepackage{newtxmath} 

\newcommand{\msun}{\ensuremath{\textup{ M}_{\odot}}}
\newcommand{\mjup}{\ensuremath{\textup{ M}_{\textsc{j}}}}
\newcommand{\mearth}{\ensuremath{\textup{ M}_{\oplus}}}
\newcommand{\dif}{\mathrm{d}}
\renewcommand{\vec}[1]{\boldsymbol{#1}}

\title[Radiative transfer in hydrodynamic simulations] 
{Efficient radiative transfer techniques in hydrodynamic simulations}
\author[Mercer et al.]
{A.~Mercer$^{1,}$\thanks{E-mail: \texttt{apmercer@uclan.ac.uk}}, 
D.~Stamatellos$^1$ \& A.~Dunhill$^1$
\\
$^1$Jeremiah Horrocks Institute for Mathematics,
Physics \& Astronomy, University of Central Lancashire, Preston, PR1 2HE, UK}

\date{Accepted XXX. Received YYY; in original form ZZZ}

\pubyear{2018}

\begin{document}
\label{firstpage}
\pagerange{\pageref{firstpage}--\pageref{lastpage}}

% ==============================================================================
% TITLE
% ==============================================================================

\maketitle

% ==============================================================================
% ABSTRACT
% ==============================================================================
\begin{abstract}
Radiative transfer is an important component of hydrodynamic simulations as it
determines the thermal properties of a physical system. It is especially
important in cases where heating and cooling regulate significant processes,
such as in the collapse of molecular clouds, the development of gravitational
instabilities in protostellar discs, disc-planet interactions, and planet
migration. We compare two approximate radiative transfer methods which
indirectly estimate optical depths within hydrodynamic simulations using two
different metrics: (i) the gravitational potential and density of the gas
(Stamatellos et al.), and (ii) the pressure scale-height (Lombardi et al.). We
find that both methods are accurate for spherical configurations e.g. in
collapsing molecular clouds and within clumps that form in protostellar discs.
However, the pressure scale-height approach is more accurate in protostellar
discs (low and high-mass discs, discs with spiral features, discs with embedded
planets). We also investigate the $\beta$-cooling approximation which is commonly
used when simulating protostellar discs, and in which the cooling time is
proportional to the orbital period of the gas. We demonstrate that the use of a
constant $\beta$ cannot capture the wide range of spatial and temporal
variations of cooling in protostellar discs, which may affect the development of
gravitational instabilities, planet migration, planet mass growth, and the orbital
properties of planets.
\end{abstract}

% ==============================================================================
% KEYWORDS
% ==============================================================================

% ==============================================================================
% INTRODUCTION
% ==============================================================================
\section{Introduction}
\label{sec:introduction}

Full 3-dimensional, wavelength dependent radiative transfer within hydrodynamic
simulations is computationally expensive \citep[e.g.][] {Harries:2015a,
Harries:2017a}. It is only typically used to post-process snapshots of
simulations to produce synthetic observations \citep[e.g.
\textsc{radmc-3d};][]{Dullemond:2012a}. However, the inclusion of radiative
transfer is important when an accurate treatment of the thermal evolution of the
system is needed.

There are various methods which efficiently include approximate radiative
transfer in hydrodynamic simulations, each with their underlying simplifying
assumptions \citep{Oxley:2003a, Whitehouse:2004a, Stamatellos:2007b,
Forgan:2009b, Young:2012a, Lombardi:2015a}. There are two main types of
approach: (i)~using the diffusion approximation
\citep[e.g.][]{Whitehouse:2004a,Boley:2006a, Commercon:2011b,Commercon:2011d}, a
method which may still be computationally expensive, or (ii) use a metric to
estimate the optical depth for each element of the fluid and hence the
heating/cooling rate \citep{Stamatellos:2007b, Forgan:2009b,
Young:2012a,Lombardi:2015a}. Another method that is used in the context of
protostellar discs is the $\beta$-cooling approximation
\citep[e.g.][]{Gammie:2001a, Rice:2003a}. This method assumes that the temporal
evolution of the specific internal energy, $u$, is inversely proportional to the
cooling time such that $\dot{u}=-u/t_{\textup{cool}}$. The cooling time is set
inversely proportional to the Keplerian frequency with a constant $\beta$, i.e.
$t_{\textup{cool}}(R)=\beta \Omega^{-1} (R)$, where $R$ is the distance from the
central star as measured on the disc midplane. This method over-simplifies the
underlying physics but comes at low computational cost.

\cite{Stamatellos:2007b} proposed a radiative transfer method which uses the
gravitational potential and the density of gas as a metric to estimate the
optical depth through which a gas element cools. This is then used to calculate
an estimated cooling rate, and, in the optically thick case, reduces to the
diffusion approximation. The method works well for roughly spherical systems and
results in an increase of computational time by less than $\sim 5 \%$. However,
\cite{Wilkins:2012a} showed that the cooling rate calculated with the
\cite{Stamatellos:2007b} method can be systematically underestimated in the
optically thick midplane of protostellar discs. Therefore, the
\cite{Stamatellos:2007b} method therefore may not be suitable to provide
accurate cooling rates in non-spherical systems. This method has been combined
with the diffusion approximation to increase accuracy in high-optical depth
regions \citep[e.g.][]{Forgan:2009b}.

\cite{Young:2012a} proposed a method, in the context of protostellar discs, that
uses the gravitational potential in the $z$ direction only, i.e. out of the disc
midplane. From this, they obtain accurate estimates (within a few tens of
percent) of column density and optical depths. However, when fragments form due
to the gravitational instability in massive discs, the \cite{Stamatellos:2007b}
gives better estimates of the cooling rates within the dense fragments, which
can be assumed to be approximately spherical.

Instead of using the gravitational potential to estimate the optical depth,
\cite{Lombardi:2015a} propose to use the pressure scale-height. This retains the
majority of the characteristics of the original \cite{Stamatellos:2007b}
method, merely employing a different metric to estimate optical depth. It is
shown to provide a much more accurate estimate of cooling rate in spherical
polytropes and protostellar discs with specified density and temperature
profiles. 

The aim of this paper is to compare how the above methods
\citep{Stamatellos:2007b, Lombardi:2015a}, behave when applied in actual
hydrodynamic simulations. We test the two methods in the context of collapsing
clouds and protostellar discs. In the case of the latter, we consider relaxed
discs, discs with spiral arms, discs with clumps, and discs with embedded
planets which carve gaps. We also examine whether the $\beta$-cooling method,
which is widely used for protostellar discs, provides a good approximation to
the thermal physics. Such tests of different methods are needed as radiative
transfer plays a critical role in many cases (e.g. disc fragmentation and gap
opening in discs with planets).

In Section \ref{sec:efficient_radiative_transfer_methods} we describe the
radiative transfer techniques in more detail. Section \ref{sec:cloud_collapse}
shows the comparison between the aforementioned methods for the collapse of
spherically-symmetric cloud. We test the behaviour of both methods for
protostellar discs in Section \ref{sec:protostellar_discs} and for discs with
embedded planets in Section \ref{sec:protostellar_discs_with_embedded_planets}.
A discussion on the performance of the $\beta$-cooling approximation is
presented in Section \ref{sec:comparison_with_the_beta_cooling_approximation}. A
comparison to demonstrate the effect on dynamical evolution from the two
radiative transfer methods discussed, as well as the $\beta$-cooling
approximation, is presented in Section \ref{sec:dynamical_evolution_comparison}.
We summarise our results in Section \ref{sec:discussion}.

% ==============================================================================
% EFFICIENT RADIATIVE TRANSFER METHODS
% ==============================================================================
\section{Efficient radiative transfer methods}
\label{sec:efficient_radiative_transfer_methods}

The radiative transfer technique ascribed to \cite{Stamatellos:2007b} is used to
determine the heating and cooling of the gas. The method incorporates the
effects from the rotational and vibrational degrees of freedom of
$\textup{H}_{2}$, the dissociation of $\textup{H}_{2}$, ice melting, dust
sublimation, bound-free, free-free, and electron scattering interactions. The
equation of state used and the effect of each assumed constituent are described
in detail in \textsection 3 of \cite{Stamatellos:2007b}. The heating/cooling
rate requires an estimate of the column density through which the
heating/cooling happens as well as the local opacity. It is expressed as
\begin{equation}
  \frac{\dif u}{\dif t} = \frac{4 \sigma_{\textsc{sb}}
  \left( T_{\textsc{bgr}}^{4} - T^{4} \right)}{\bar{\Sigma}^{2}
  \bar{\kappa}_{\textsc{r}}\left(\rho, T \right) +
  \kappa_{\textsc{p}}^{-1}\left(\rho, T \right)},
\label{eqn:coolingRate}
\end{equation}
where $\sigma_{\textsc{sb}}$ is the Stefan-Boltzmann constant,
$T_{\textsc{bgr}}$ is the pseudo-background temperature below which the gas
cannot cool radiatively, $\bar{\Sigma}$ is the mass-weighted mean column
density, and $\bar{\kappa}_{\textsc{r}}$ and $\kappa_{\textsc{p}}$ are the
Rosseland- and Planck-mean opacities, respectively. In the original
\cite{Stamatellos:2007b} method, the estimated mass-weighted column density is
found via the local density $\rho$ and gravitational potential $\psi$ such that
\begin{equation}
  \bar{\Sigma} = \zeta \left( \frac{-\psi \rho}
  {4\pi G}\right)^{1/2},
\label{eqn:mean_weighted_cd}
\end{equation}
where $\zeta = 0.372$ is a dimensionless coefficient with a weak dependence on
polytropic index (set to $n=1.5$). Particles are assumed to be surrounded by a
pseudo-cloud represented by a polytrope. A particle heats or cools according to
the characteristic optical depth of its pseudo-cloud (wherein the particle can
be located at any position to account for non-spherical local geometry). The
optical depth can be found via
\begin{equation}
 {\tau} = \bar{\Sigma} \bar{\kappa}.
\label{eqn:mean_weighted_tau}
\end{equation}
When considering the collapse of a 1\msun~spherical cloud of gas, the method has
been shown to produce similar results to the simulations of
\cite{Masunaga:2000a}, which is a 1-D hydrodynamic simulation where the
radiative transfer is treated accurately \citep{Stamatellos:2007b}. 

\cite{Lombardi:2015a} argue that the use of the gravitational potential as a
metric overestimates column densities and optical depths in non-spherical
configurations such as discs. Instead, they propose the use of pressure
scale-height as a metric for calculating the optical depth. This is because the
pressure gradient is typically oriented in the direction in which the optical
depth increases most rapidly, i.e. approximately perpendicular to the disc
midplane. The \cite{Lombardi:2015a} form for the estimated mass-weighted column
density is
\begin{equation}
  \bar{\Sigma} = \zeta' \frac{P}{\left| \vec{a}_{h}\right|},
\label{eqn:lombardi_sigma}
\end{equation}
where $\zeta' = 1.014$ is a dimensionless coefficient. $P$ is the pressure of
the gas and $\vec{a}_{h}$ the hydrodynamical acceleration (i.e. the acceleration
without any gravitational nor viscous contribution). This quantity can be
expressed in terms of the pressure gradient such that
\begin{equation}
  \vec{a}_{h} = \frac{-\nabla P}{\rho}.
\label{eqn:hydro_acc}
\end{equation}
For either method, the required quantities are readily available in any
hydrodynamic method.

For a given particle density and temperature, a density-sorted look-up table can
be used to find: specific internal energy; mean-molecular mass; mass-weighted
optical depth; Rosseland- and Planck-mean optical depths; ratio of specific heat
capacities; and the first adiabatic index. This removes the requirement of
calculating computationally-expensive integrals on-the-fly (see \textsection 2.2
of \cite{Stamatellos:2007b}). 

We note that although the above methods have been devised for Smooth Particle
Hydrodynamics \citep{Gingold:1977a, Lucy:1977a}, they can be applied to
grid-based \citep[e.g.][]{Fryxell:2000a} and meshless techniques
\citep{Lanson:2008a, Gaburov:2011a, Hopkins:2015b}. 

% ==============================================================================
% CLOUD COLLAPSE
% ==============================================================================
\section{Cloud collapse}
\label{sec:cloud_collapse}

We utilise the Graphical Astrophysics code for N-body Dynamics and Lagrangian
Fluids \citep[\textsc{gandalf},][]{Hubber:2018a} to perform simulations of a
collapsing molecular cloud, using the \cite{Stamatellos:2007b} and
\cite{Lombardi:2015a} methods of estimating optical depths. The cloud is
initially static, has a mass of $1.5\msun$ and is isothermal with a temperature
5~K. The cloud is represented by $N \approx 2 \times 10^{6}$ SPH particles
distributed such that the density profile of the cloud is uniform across its
radius $R_{\textup{cloud}} = 10^{4}$ AU.

\begin{figure}
  \begin{center}
    \includegraphics[width = 0.5\textwidth, trim = 0cm 0cm 0cm 0cm,
    clip=true]{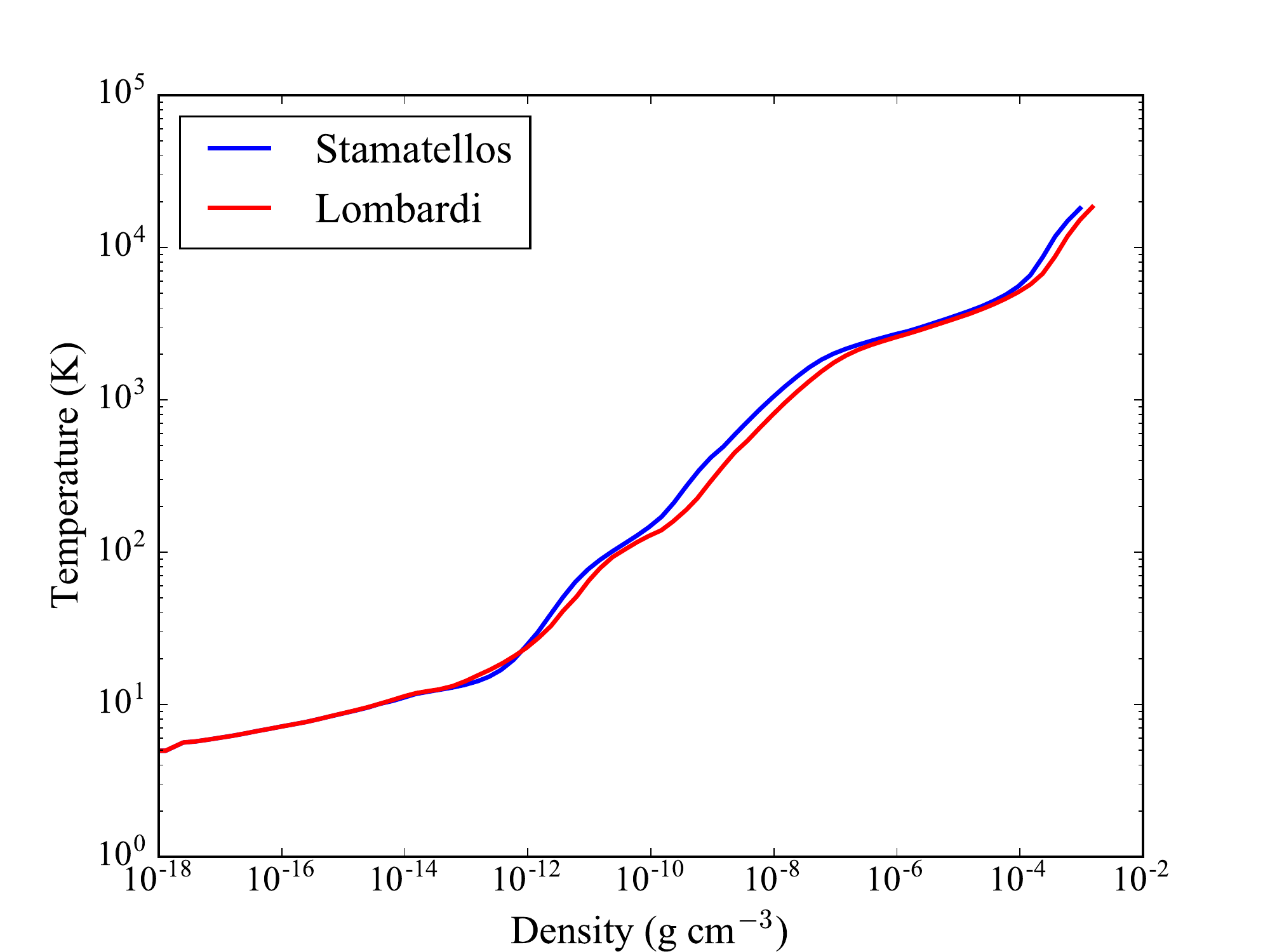}
    \caption
    {
      The evolution of central temperature as a function of central density for
      the collapse of an initially isothermal, non-rotating, $1.5 \msun$ cloud
      with a radius of $10^{4}$ AU. The radiative transfer methods of
      \protect\cite{Stamatellos:2007b} and \protect\cite{Lombardi:2015a} are in
      good agreement.
    }
    \label{fig:cloud_collapse_test}
  \end{center}
\end{figure}

Figure \ref{fig:cloud_collapse_test} shows the evolution of the central density
and temperature for the two methods of estimating optical depths. Initially, the
cloud collapses almost isothermally and the core temperature increases slowly
with increasing density. The core temperature starts to increase rapidly as the
cloud becomes optically thick ($\rho\sim10^{-13} \textup{ g cm}^{-3}$). At $\sim
100$~K the rotational degrees of freedom of molecular hydrogen are excited and
the temperature increases at a slower rate as the gravitational energy is
diverted away from heating the cloud. The increasing temperature leads to
increased thermal pressure that is able to slow down the collapse and the first
hydrostatic core forms \citep{Larson:1969a, Masunaga:2000a, Whitehouse:2006a,
Stamatellos:2007b}. The first core contracts and heats slowly to $\sim 2000$~K
at which point hydrogen begins to dissociate. This results in the second
collapse and the formation of the second hydrostatic core (the protostar). 

The \cite{Lombardi:2015a} method gives similar results regarding the central
density and temperature of the cloud with the \cite{Stamatellos:2007b} method,
which is turn compares very well with the \cite{Masunaga:2000a} method,
indicating that both methods work reasonably well for spherical geometries. The
second collapse in the case of the \cite{Stamatellos:2007b} method is delayed by
$\sim 100$~yr, which may arise due to a slight over-estimate in optical depth
and thus less efficient cooling, as can be seen from the slightly higher
temperatures calculated by this method (see Fig.~\ref{fig:cloud_collapse_test}).

% FIGURE : CLOUD COLLAPSE COMPARISON -------------------------------------------

% ==============================================================================
% PROTOSTELLAR DISCS
% ==============================================================================
\section{Protostellar discs}
\label{sec:protostellar_discs}

Protostellar discs form due to the turbulence and/or initial rotation of their
progenitor molecular clouds. Their study is important as they are the birthplace
of planets, which can form either through core accretion
\citep[e.g][]{Safronov:1969a, Lissauer:1993a}, or by  gravitational
fragmentation of discs \citep{Whitworth:2006a, Stamatellos:2007c,
Stamatellos:2009a, Kratter:2010a, Zhu:2012a}. Massive protostellar discs
fragment if two conditions are met: (i) They are gravitationally unstable i.e. 
\begin{equation}
  Q \equiv \frac{\kappa c_{s}}{\pi G \Sigma} < Q_{\textup{crit}},
\label{eqn:toomre}
\end{equation}
where $Q$ is the Toomre parameter \citep{Toomre:1964a}, $\kappa$ is the
epicyclic frequency, $c_{s}$ is the local sound speed and $\Sigma$ is the disc
surface density. The value of $ Q_{\textup{crit}}$ is on the order of unity.
(ii)~They cool sufficiently fast, i.e.
$t_{\textsc{cool}}<(0.5-2)t_{\textsc{orb}}$, where $t_{\textsc{orb}}$ is the
local orbital period \citep{Gammie:2001a,Johnson:2003a, Rice:2003a, Rice:2005a}.
Both requirements are dependent on the thermal properties of the disc, and so it
is important that the cooling rate and the disc temperature are accurately
calculated with the employed radiative transfer method.

Here, we present comparisons of estimated optical depth and cooling rate
obtained via the \cite{Stamatellos:2007b} and \cite{Lombardi:2015a} radiative
transfer methods. In Section~\ref{sec:methodology} we present our comparison
methodology. Section \ref{sub:relaxed_low_mass_disc} considers a low-mass
relaxed disc. Section \ref{sub:high_mass_disc} considers a high-mass
disc which fragments forming spiral arms (Section
\ref{sub:disc_with_spiral_arms}) and eventually gravitationally bound clumps
(Section \ref{sub:disc_with_clumps}).

% ==============================================================================
% DISC METHODOLOGY
% ==============================================================================
\subsection{Methodology}
\label{sec:methodology}

We use the Graphical Astrophysics code for N-body Dynamics and Lagrangian Fluids
\citep[\textsc{gandalf},][]{Hubber:2018a} to perform simulations protostellar
discs (\textsection \ref{sec:protostellar_discs}) and protostellar discs with
embedded planets (\textsection
\ref{sec:protostellar_discs_with_embedded_planets}). From these simulations we
select snapshots for which we compare the behaviour of the
\cite{Stamatellos:2007b} and \cite{Lombardi:2015a} radiative transfer methods.

The estimated column density for both the gravitational potential and pressure
scale-height metrics, $\bar{\Sigma} \equiv \Sigma_{\textup{est}}$, is found by
post-processing a snapshot of the \textsc{gandalf} hydrodynamic simulation. The
corresponding estimated optical depth is $\bar{\tau} \equiv \tau_{\textup{est}}
= \Sigma_{\textup{est}} \bar{\kappa}_{\textsc{r}}$, where
$\bar{\kappa}_{\textsc{r}}$ is the mass-weighted opacity for each method (note
that this is slightly different for the two methods, see \cite{Lombardi:2015a}).
The column density and optical depth are calculated for each particle in the
simulation. We emphasise that we calculate the optical depths and cooling rates
for the same snapshots for both methods, i.e. using the same density and
temperature disc configurations. We provide azimuthally-averaged radial
profiles of the optical depth and cooling rates at the disc midplane (defined
such as $|z|<0.5$~AU) and also vertical to the disc midplane profiles of the
same quantities. We also calculate the actual values of column density and
optical depth by integrating from the gas element which we consider, to the disc
surface along the $z$-axis (perpendicular to the disc midplane) such that
$\Sigma_{\textup{actual}} = \int \rho ~ \mathrm{d}z$ and $\tau_{\textup{actual}}
= \int \kappa\left(\rho, T\right) \rho ~ \mathrm{d}z$.

The estimated cooling-rate per unit mass can then be found via Equation
\ref{eqn:coolingRate}. We normalise this with respect to ${4
\sigma_{\textsc{sb}} \left( T^{4} - T_{\textsc{bgr}}^{4}\right)}$ such that we
define the quantity
\begin{equation}
  \dot{u}_{\textup{est}} \equiv -\left.\frac{\mathrm{d}u}{\mathrm{d} t}
  \right|_{\textup{est}}
  \frac{1}{4 \sigma_{\textsc{sb}} \left( T^{4} - T_{\textsc{bgr}}^{4}\right)} =
  \frac{1}{\bar{\Sigma}^{2}\bar{\kappa}_{\textsc{r}} +
  \kappa_{\textsc{p}}^{-1}}
\label{eqn:pseudo_cooling_rate}
\end{equation}
to represent the estimated cooling-rate per unit mass. We compare this with the
actual cooling-rate per unit mass which is calculated using the actual optical
depth and column density, hence
\begin{equation}
  \dot{u}_{\textup{actual}} \equiv -\left.\frac{\mathrm{d}u}{\mathrm{d} t}
  \right|_{\textup{actual}}
  \frac{1}{4 \sigma_{\textsc{sb}} \left( T^{4} - T_{\textsc{bgr}}^{4}\right)} =
  \frac{1}{\Sigma\left(\tau_{\textsc{r}} + \tau_{\textsc{p}}^{-1}\right)},
\label{eqn:real_cooling_rate}
\end{equation}
where $\tau_{\textsc{r}}$ and $\tau_{\textsc{p}}$ are the optical depths
calculated using the Rosseland-mean and Planck-mean opacities, respectively
(which in many cases are assumed to be the same). We note that the above
equation is itself an approximation to the diffusion approximation
\citep{Mihalas:1970a} in which the radiative flux is 
\begin{equation}
  F = -\frac{4}{3\kappa_{\textsc{r}}\rho}
      \nabla\left(\sigma_{\textsc{sb}}T^{4}\right).
\label{eqn:diffusion_approximation}
\end{equation}
From this, we obtain the cooling rate per unit mass which is
\begin{equation}
  \dot{u} = \frac{1}{\rho} \nabla \cdot F \approx 
  \frac{\sigma_{\textsc{sb}}T^{4}}{\kappa_{\textsc{r}} \Sigma^{2}} \approx 
  \frac{\sigma_{\textsc{sb}}T^{4}}{\tau_{\textsc{r}} \Sigma},\,
\end{equation}
and has the same form of Equation~\ref{eqn:real_cooling_rate} in the optically
thick limit.

% ------------------------------------------------------------------------------
% RELAXED LOW-MASS DISC
% ------------------------------------------------------------------------------
\subsection{Relaxed low-mass disc}
\label{sub:relaxed_low_mass_disc}

We simulate a protostellar disc with a mass of $0.01 \msun$ around a $1 \msun$
protostar. $N \approx 2 \times 10^{6}$ SPH particles are distributed between
radii of 5 and 100 AU such that the initial column density and temperature
profiles follow $\Sigma(R) \propto R^{-1}$ and $T(R) \propto R^{-1/2}$,
respectively. The temperature at 1 AU from the central star is $T_{0} = 250$~K.
The disc is heated by an ambient radiation field of 10 K.

\begin{figure*}
    \begin{centering}
    \subfloat{\includegraphics[width =0.465\textwidth, trim = 0cm 0cm 0cm 0cm,
    clip=true]{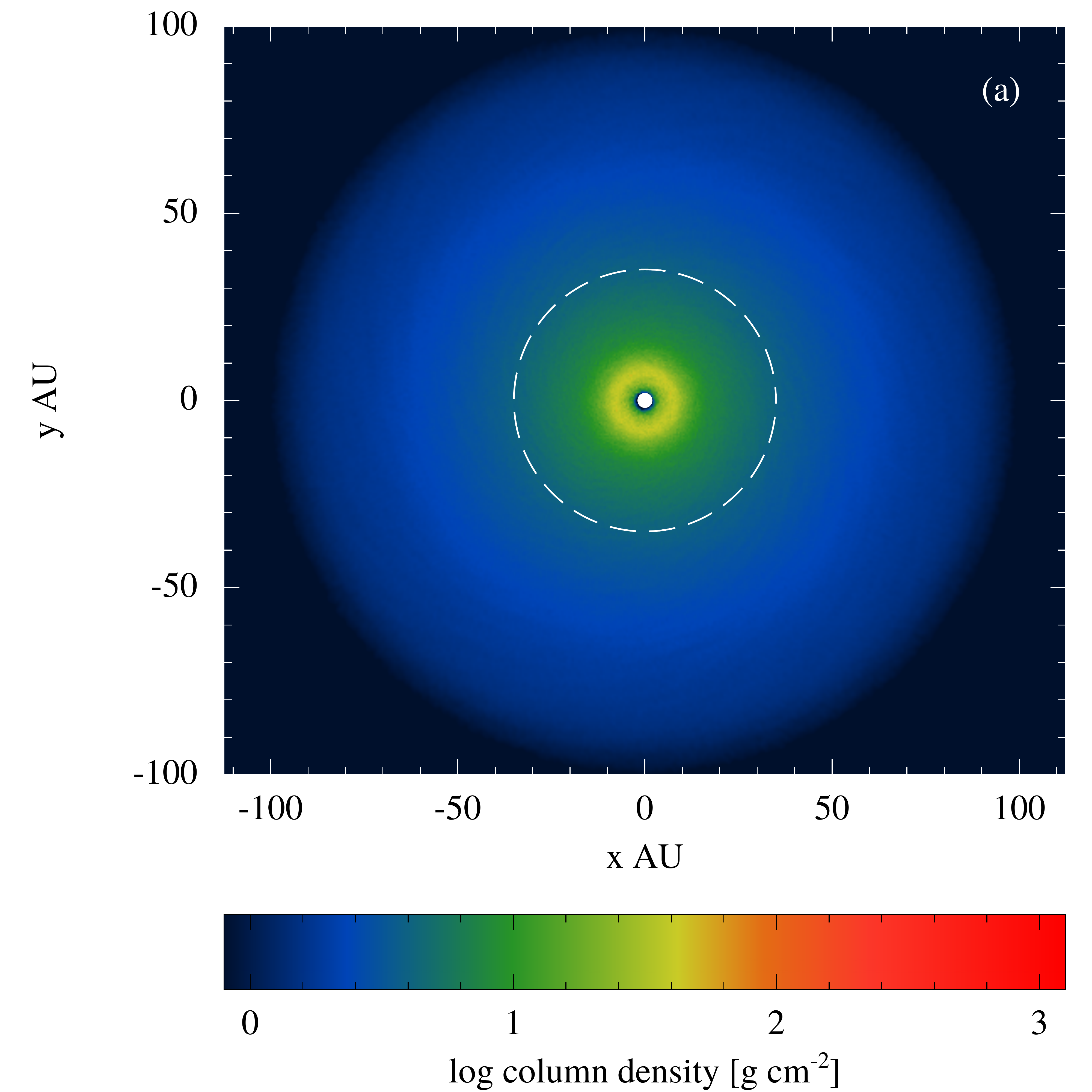}} \\
    \subfloat{\includegraphics[width =0.465\textwidth, trim = 0cm 0cm 0cm 0cm,
    clip=true]{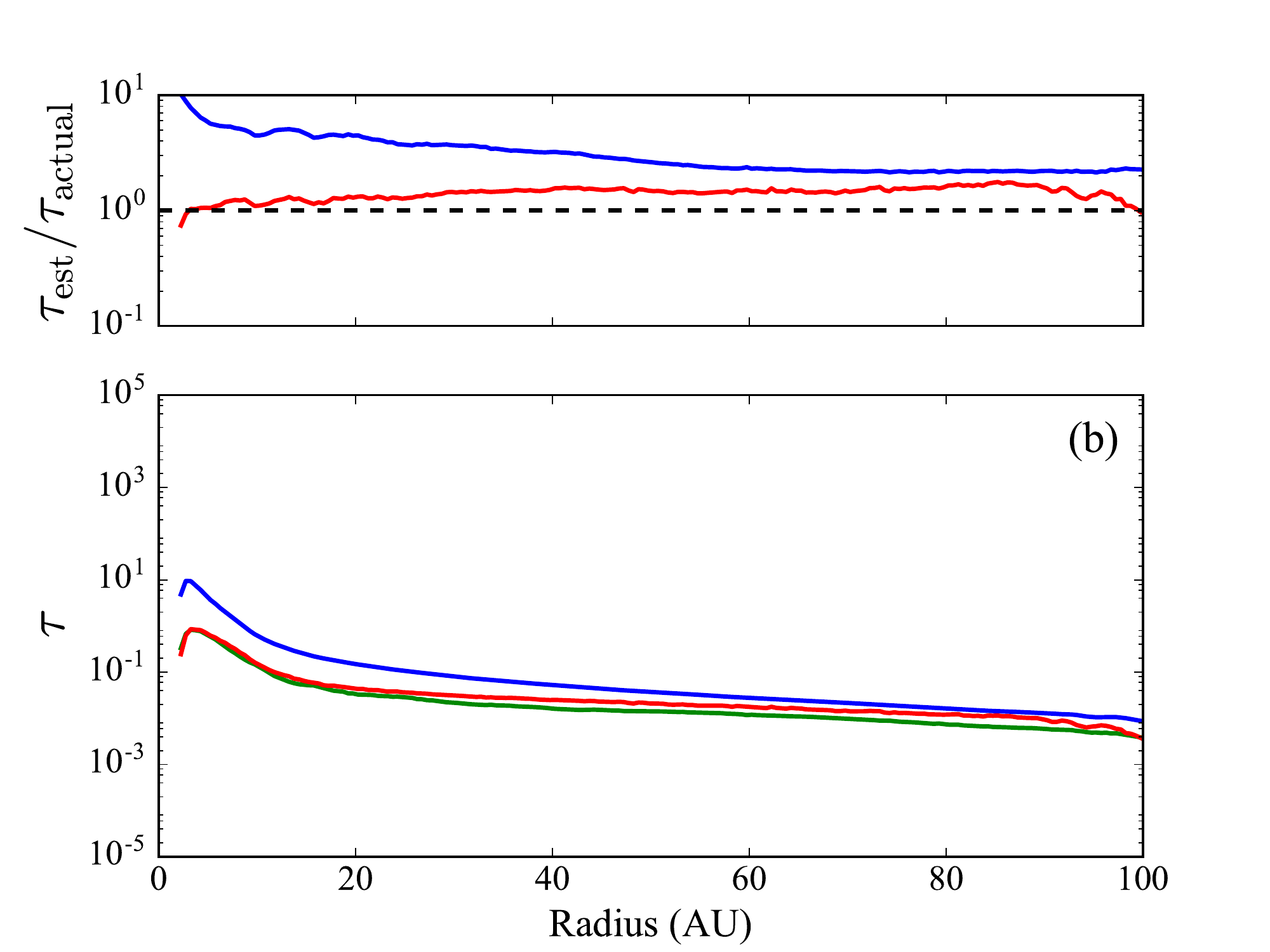}}
    \subfloat{\includegraphics[width =0.465\textwidth, trim = 0cm 0cm 0cm 0cm,
    clip=true]{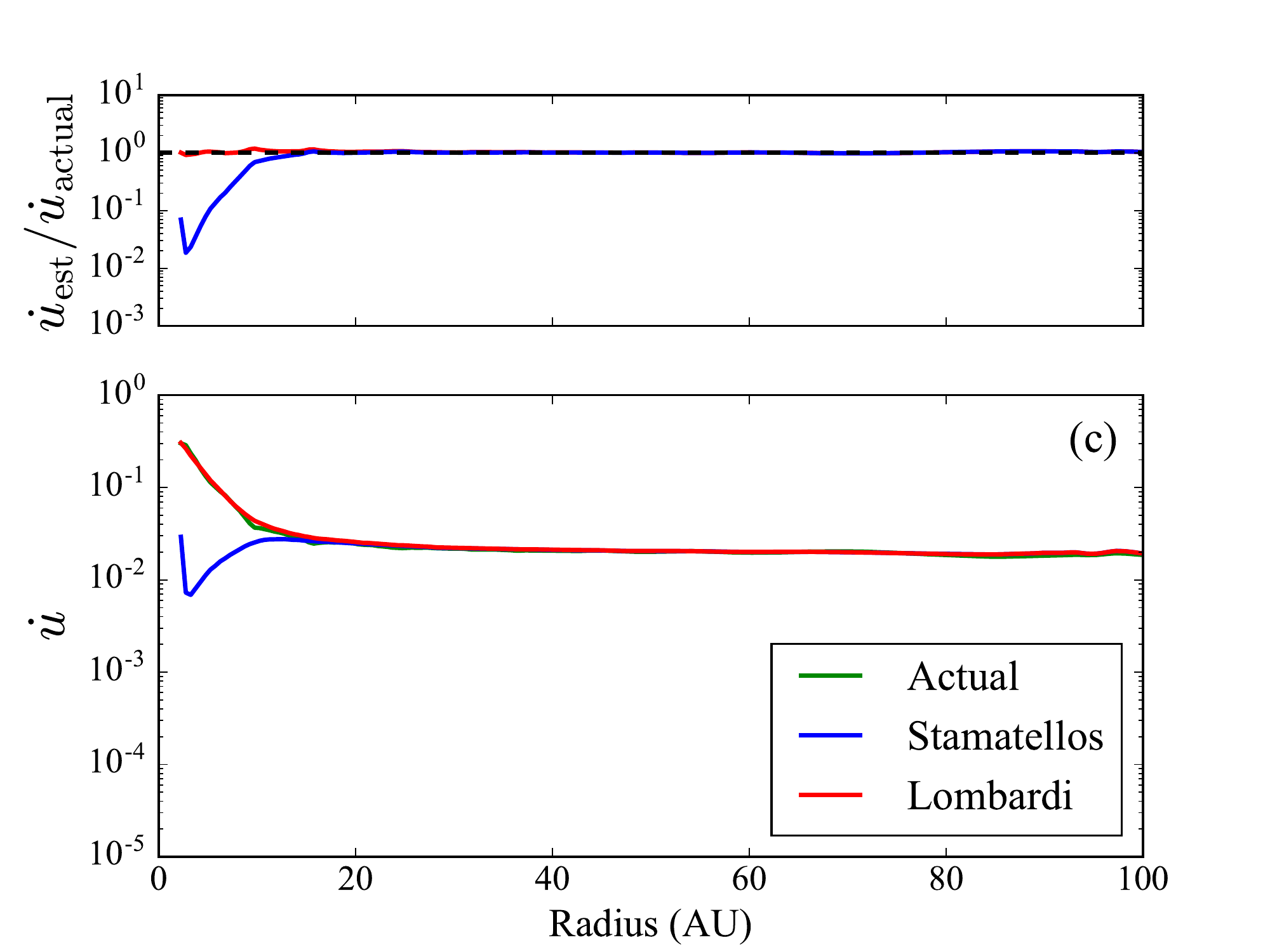}} \\
    \subfloat{\includegraphics[width =0.465\textwidth, trim = 0cm 0cm 0cm 0cm,
    clip=true]{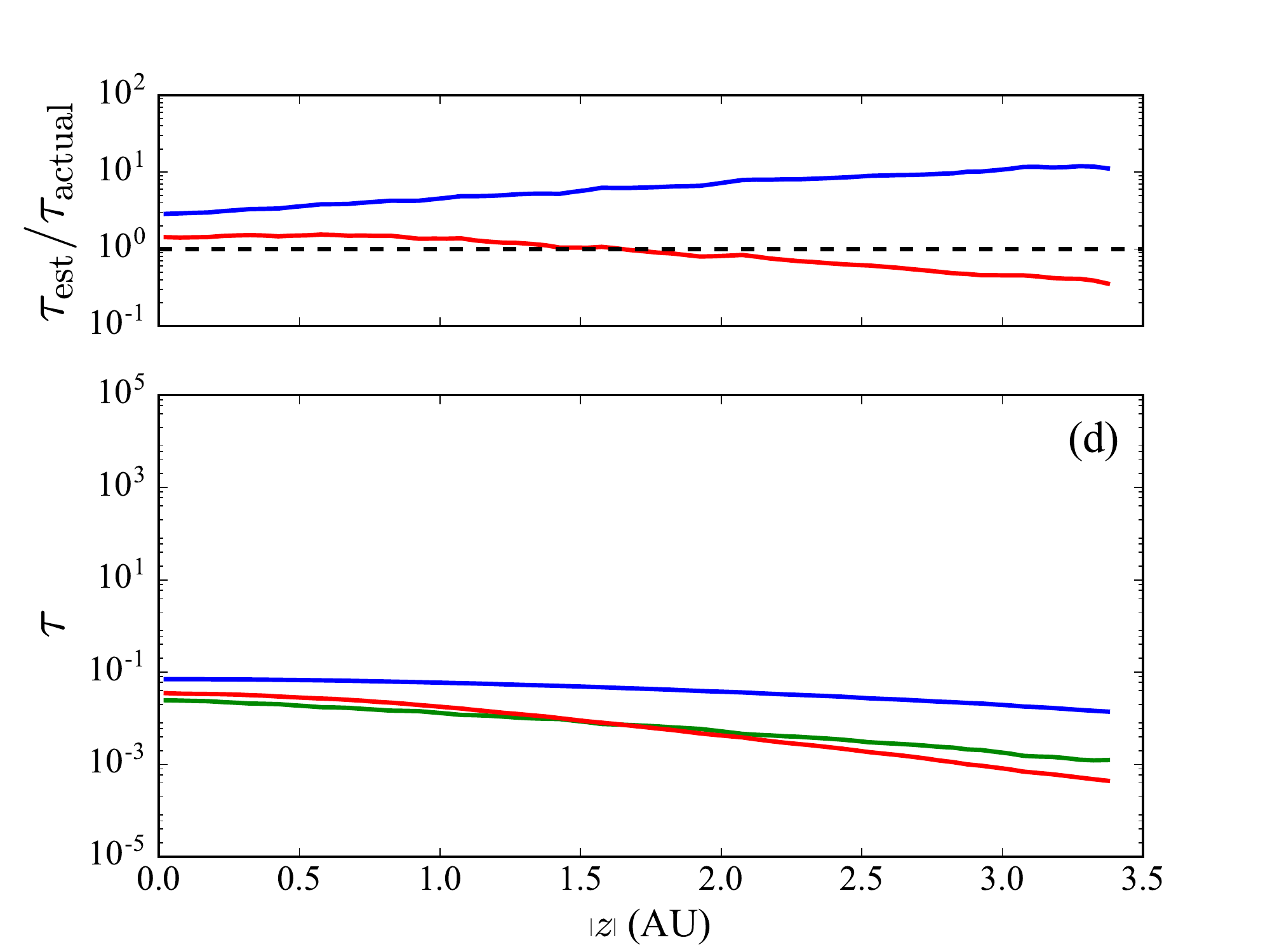}}
    \subfloat{\includegraphics[width =0.465\textwidth, trim = 0cm 0cm 0cm 0cm,
    clip=true]{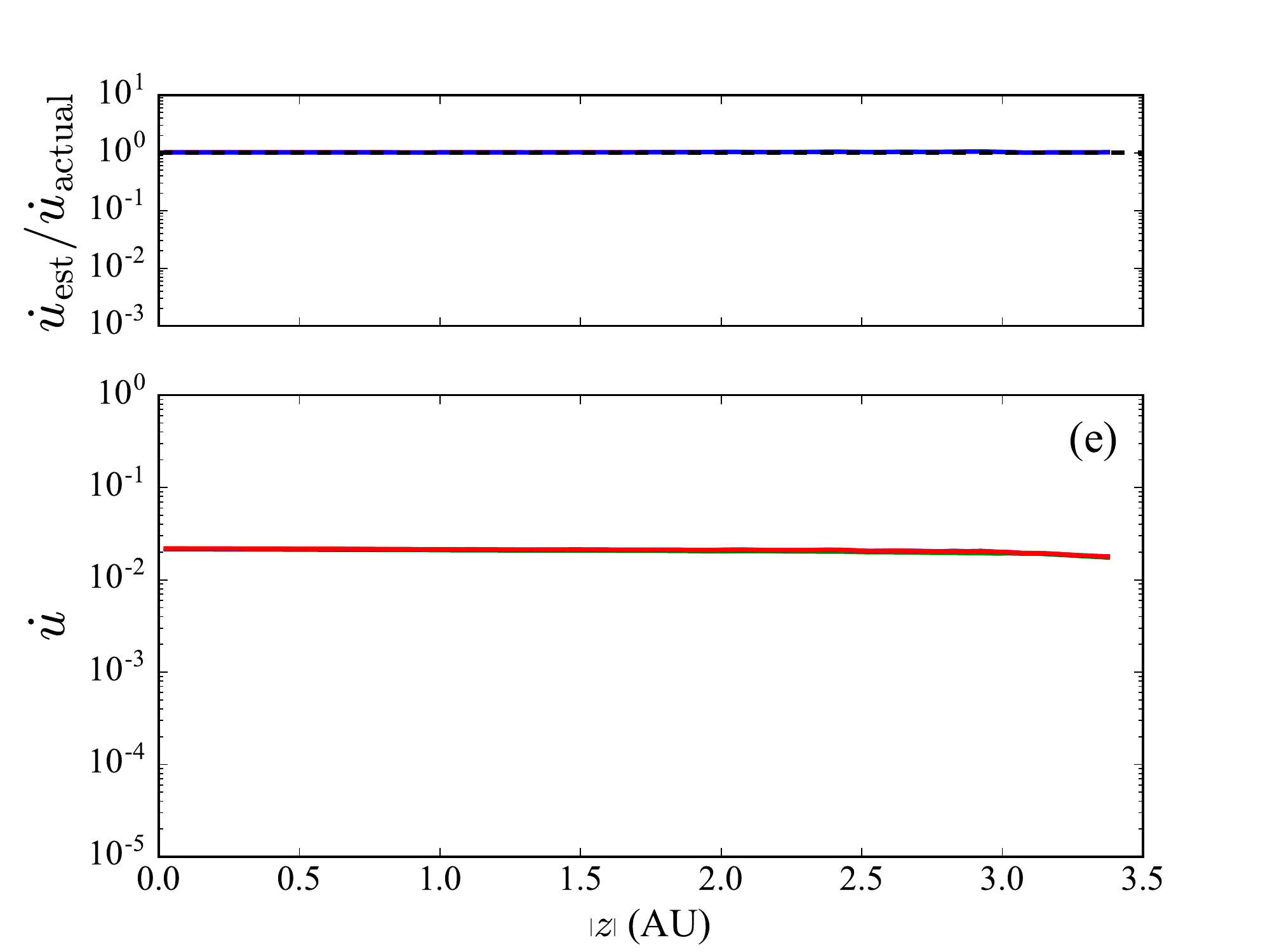}}
    \caption
    {
      A low-mass disc which has evolved for a few outer orbital periods and has
      reached a steady-state. Panel (a): a column density snapshot where the
      dashed white line represents the radius at which we perform an analysis
      perpendicular to the disc midplane. Panels (b) and (c): comparisons of
      azimuthally-averaged optical depth and cooling rate at the disc midplane
      ($|z|<0.5$~AU). Panels (d) and (e): azimuthally-averaged optical depth and
      cooling rate perpendicular to the disc midplane for a radial annulus of
      $34-36$~AU. The upper plots in panels (b-e) show the ratio between
      estimated and actual values. The black dashed lines represent equality.
      The disc is optically thin, and as such, both methods give good
      estimates of the cooling rate. The Stamatellos method generally
      overestimates the optical depth at the disc midplane, especially in the
      inner disc region, consequently underestimating the cooling rate.
    }
    \label{fig:lowmass_comparison}
    \end{centering}
\end{figure*}

A steady-state is reached after a few outer orbital periods, shown in Figure
\ref{fig:lowmass_comparison}a. The disc is optically thin, thus both the
Stamatellos and Lombardi methods provide accurate cooling rate estimates (see
Figure \ref{fig:lowmass_comparison}b). However, the Stamatellos method generally
overestimates the optical depth, especially in the inner disc, consequently
underestimating the cooling rate. We also take an annulus of the disc between 34
and 36 AU and show the azimuthally-averaged vertical profiles of optical depth
and cooling rate (Figure \ref{fig:lowmass_comparison}d, e). The cooling rate
from the disc midplane to the surface is accurately estimated as the region is
optically thin. In this regime, the optical depth is not important for
calculating the cooling rate (see Equation~\ref{eqn:coolingRate}).

% FIGURE : LOW-MASS DISC -------------------------------------------------------

% ------------------------------------------------------------------------------
% HIGH-MASS DISC
% ------------------------------------------------------------------------------
\subsection{High-mass disc}
\label{sub:high_mass_disc}

We simulate a massive protostellar disc which develops spiral features,
undergoes fragmentation, forming dense, gravitationally-bound clumps. The disc
has an initial mass of $0.2 \msun$ and attends a $0.8 \msun$ protostar. $N
\approx 2 \times 10^{6}$ SPH particles are distributed between radii of 5 and
100 AU such that the initial column density and temperature profiles follow
$\Sigma(R) \propto R^{-1}$ and $T(R) \propto R^{-1/2}$, respectively. The
temperature at 1 AU from the central star is $T_{0} = 250$~K. The disc is heated
by an ambient radiation field of 10~K.

\begin{figure*}
    \begin{centering}
    \subfloat{\includegraphics[width =0.465\textwidth, trim = 0cm 0cm 0cm 0cm,
    clip=true]{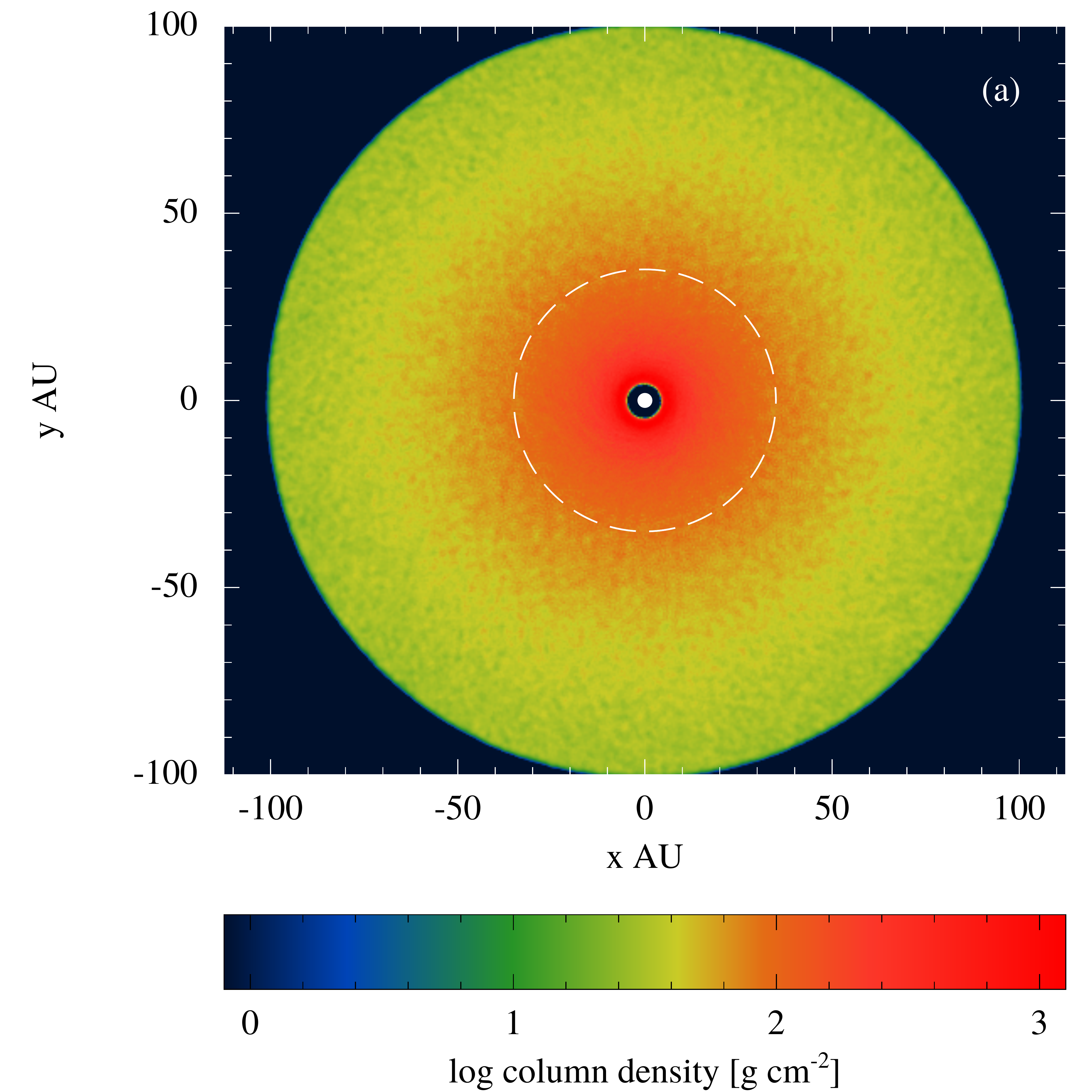}} \\
    \subfloat{\includegraphics[width =0.465\textwidth, trim = 0cm 0cm 0cm 0cm,
    clip=true]{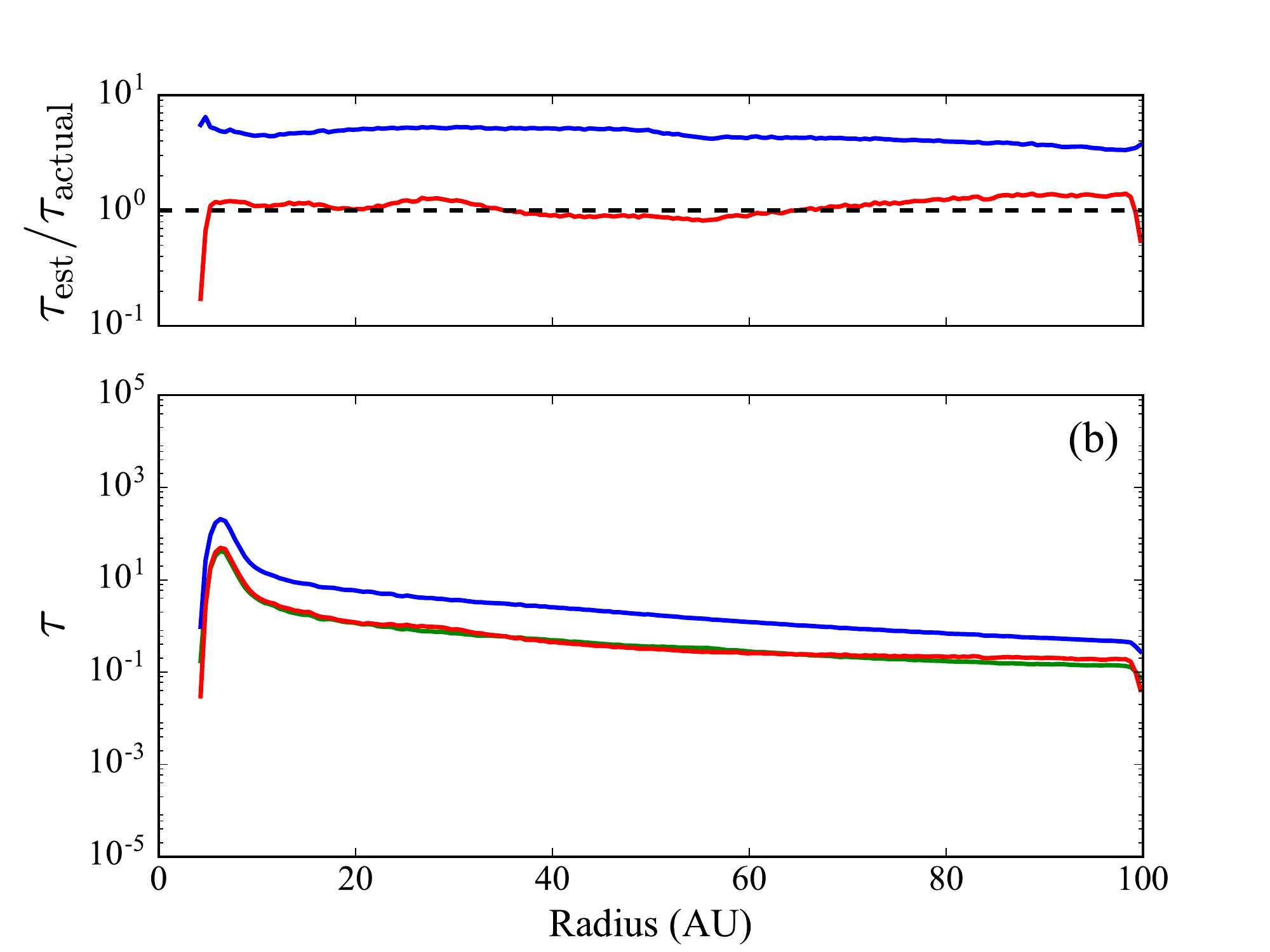}}
    \subfloat{\includegraphics[width =0.465\textwidth, trim = 0cm 0cm 0cm 0cm,
    clip=true]{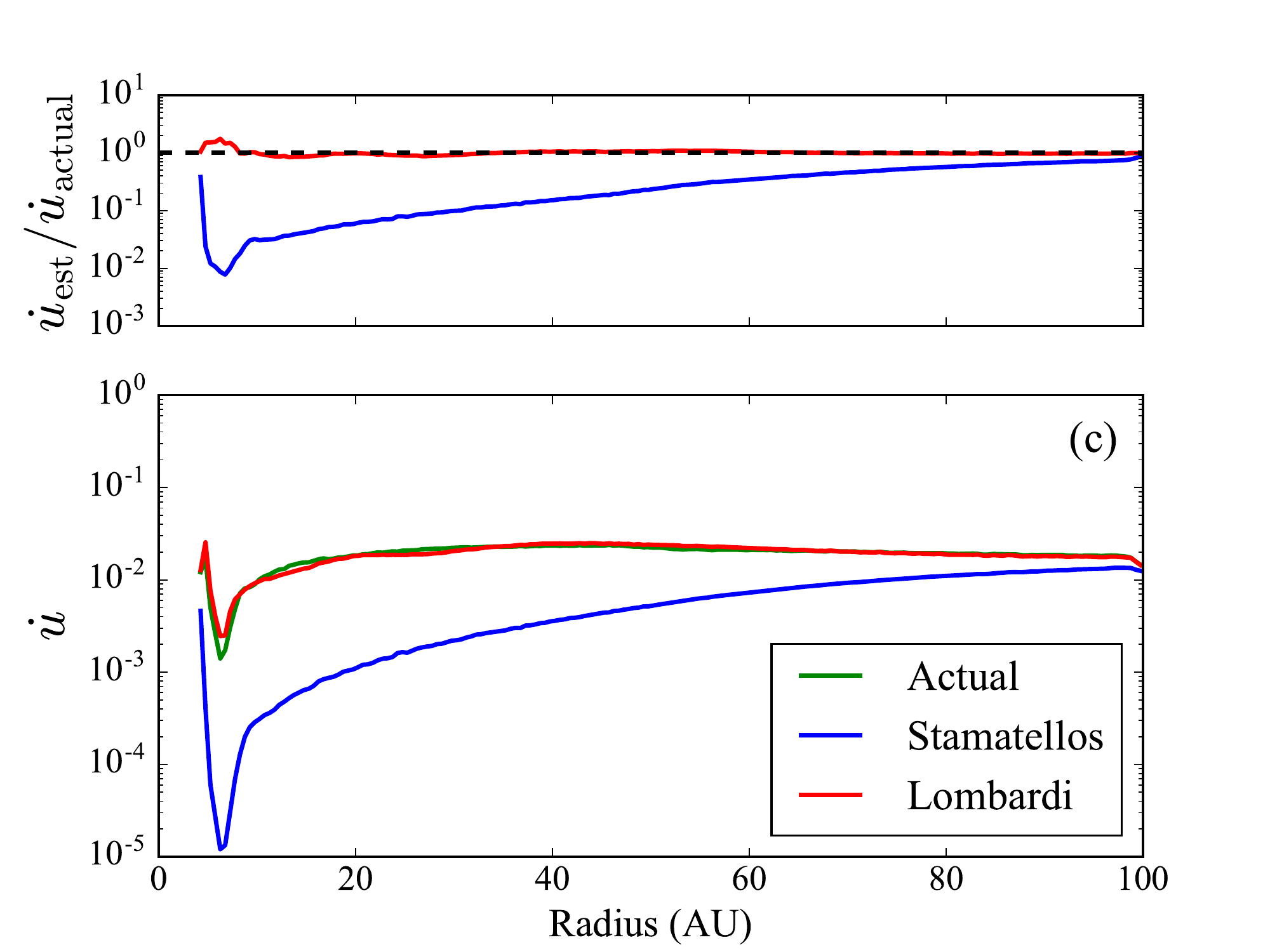}} \\
    \subfloat{\includegraphics[width =0.465\textwidth, trim = 0cm 0cm 0cm 0cm,
    clip=true]{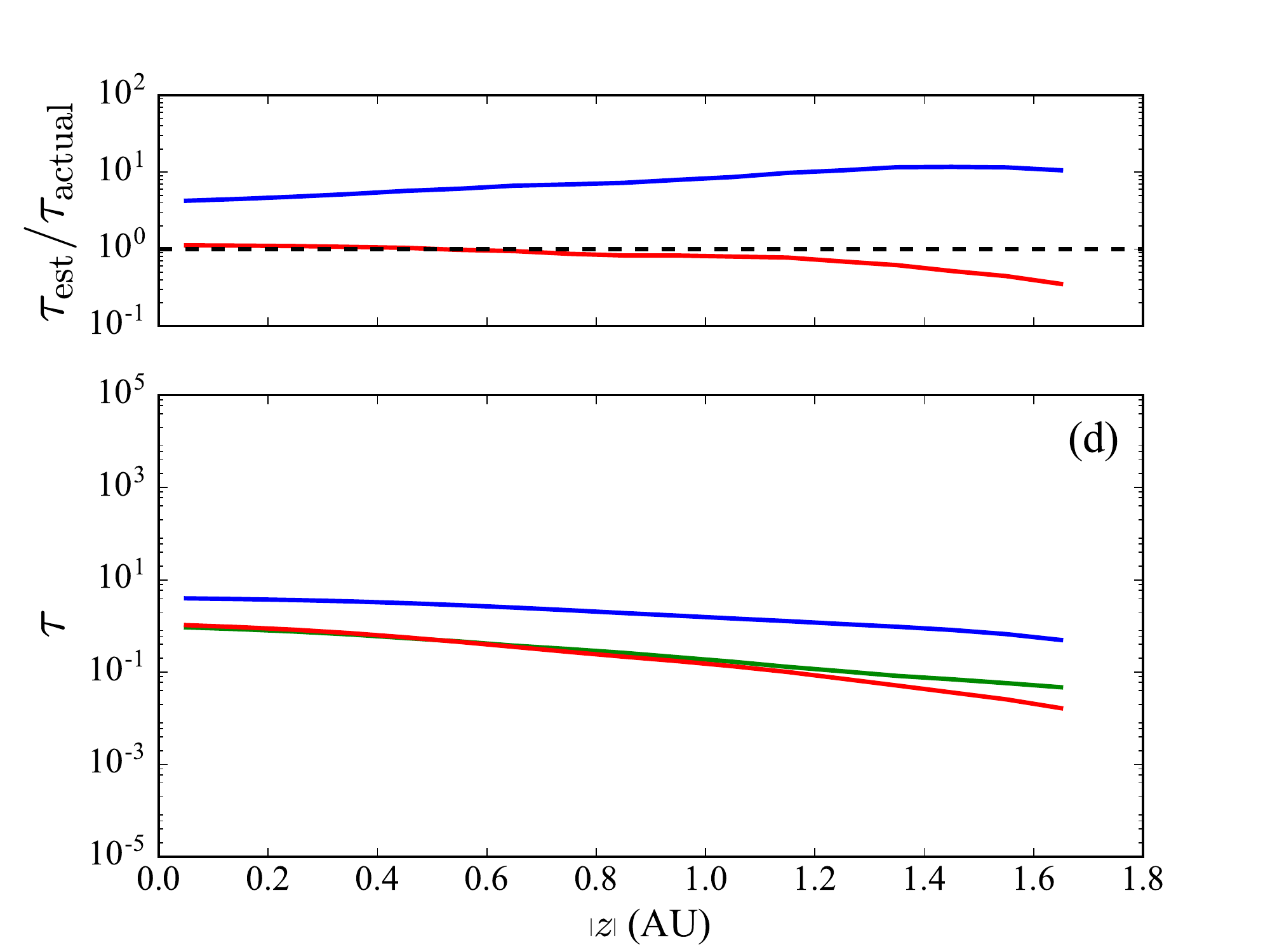}}
    \subfloat{\includegraphics[width =0.465\textwidth, trim = 0cm 0cm 0cm 0cm,
    clip=true]{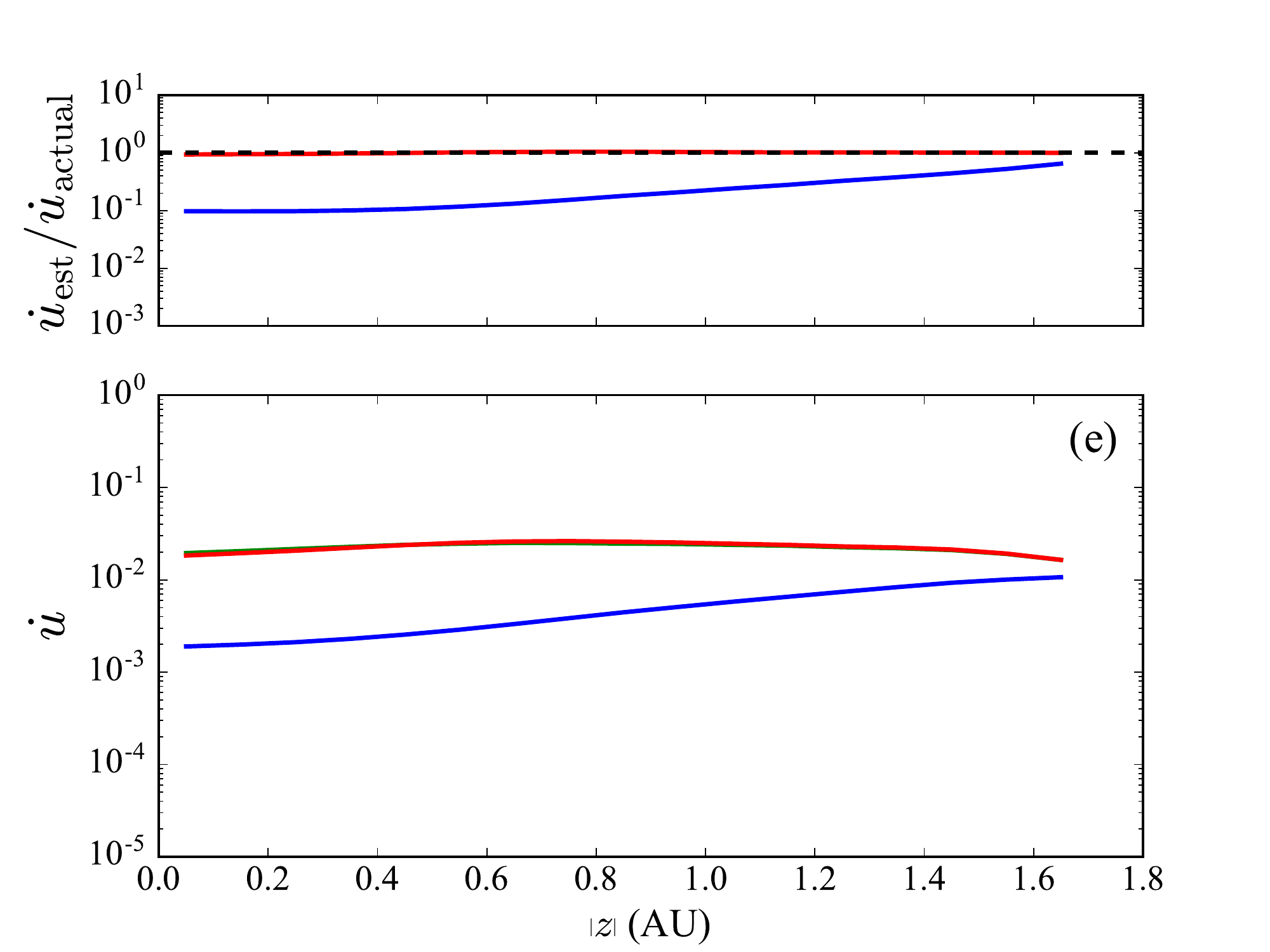}}
    \caption
    {
      A high-mass disc which has not yet undergone significant evolution. Panel
      (a): a column density snapshot where the dashed white line represents the
      radius at which we perform an analysis perpendicular to the disc midplane.
      Panels (b) and (c): comparisons of azimuthally-averaged optical depth and
      cooling rate at the disc midplane. Panels (d) and (e):
      azimuthally-averaged optical depth and cooling rate perpendicular to the
      disc midplane for a radial annulus of $34-36$~AU. The upper plots in
      panels (b-e) show the ratio between estimated and actual values. The black
      dashed lines represent equality. The Stamatellos method overestimates the
      optical depth at the disc midplane by a factor $\sim 5$ at all disc radii,
      but the Lombardi method yields a more accurate estimate. This is reflected
      in the cooling rate. Similar results are found when considering the
      optical depth and cooling profiles perpendicular to the disc midplane
      (d-e).
    }
    \label{fig:highmass_comparison}
    \end{centering}
\end{figure*}

Figure \ref{fig:highmass_comparison}a shows the column density of the disc
before any significant dynamical evolution occurs. The disc midplane is
optically thick (out to a radius of $\sim 30$~AU), but the optical depth does
not drop below $\tau = 0.1$ further out (Figure \ref{fig:highmass_comparison}b).
The Stamatellos method overestimates the optical depth by a factor of a few
throughout the disc. The Lombardi method yields a better estimate for both the
optical depth and the cooling rate. Similar results are found when considering
the vertical profiles of these quantities in a radial annulus between 34 and 36
AU (Figure \ref{fig:highmass_comparison}d, e).

% FIGURE : HIGH-MASS SNAPSHOT COMPARISON ---------------------------------------

% ------------------------------------------------------------------------------
% DISC WITH SPIRAL ARMS
% ------------------------------------------------------------------------------
\subsection{High-mass disc with spiral arms}
\label{sub:disc_with_spiral_arms}

\begin{figure*}
    \begin{centering}
    \subfloat{\includegraphics[width =0.465\textwidth, trim = 0cm 0cm 0cm 0cm,
    clip=true]{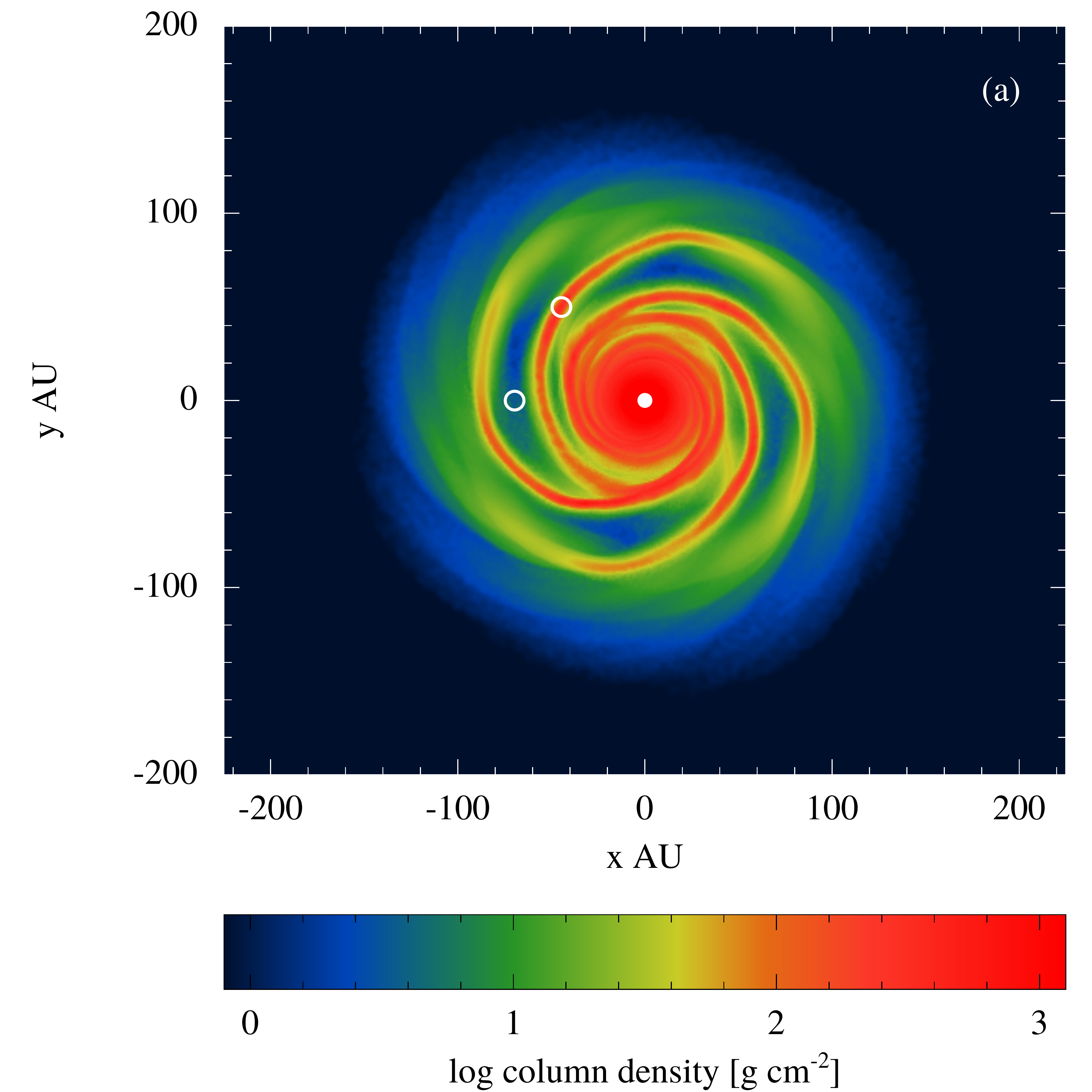}} \\
    \subfloat{\includegraphics[width =0.465\textwidth, trim = 0cm 0cm 0cm 0cm,
    clip=true]{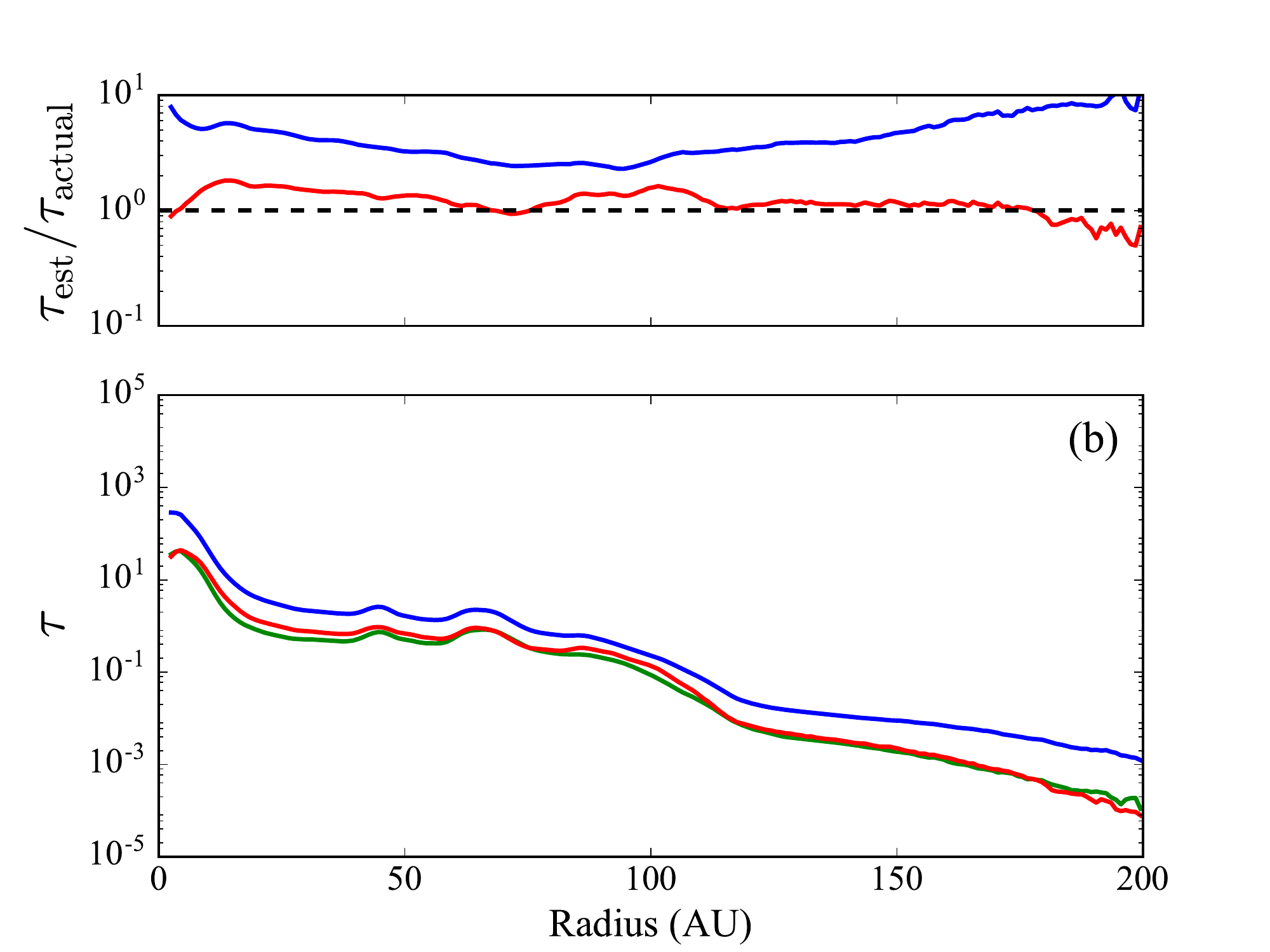}}
    \subfloat{\includegraphics[width =0.465\textwidth, trim = 0cm 0cm 0cm 0cm,
    clip=true]{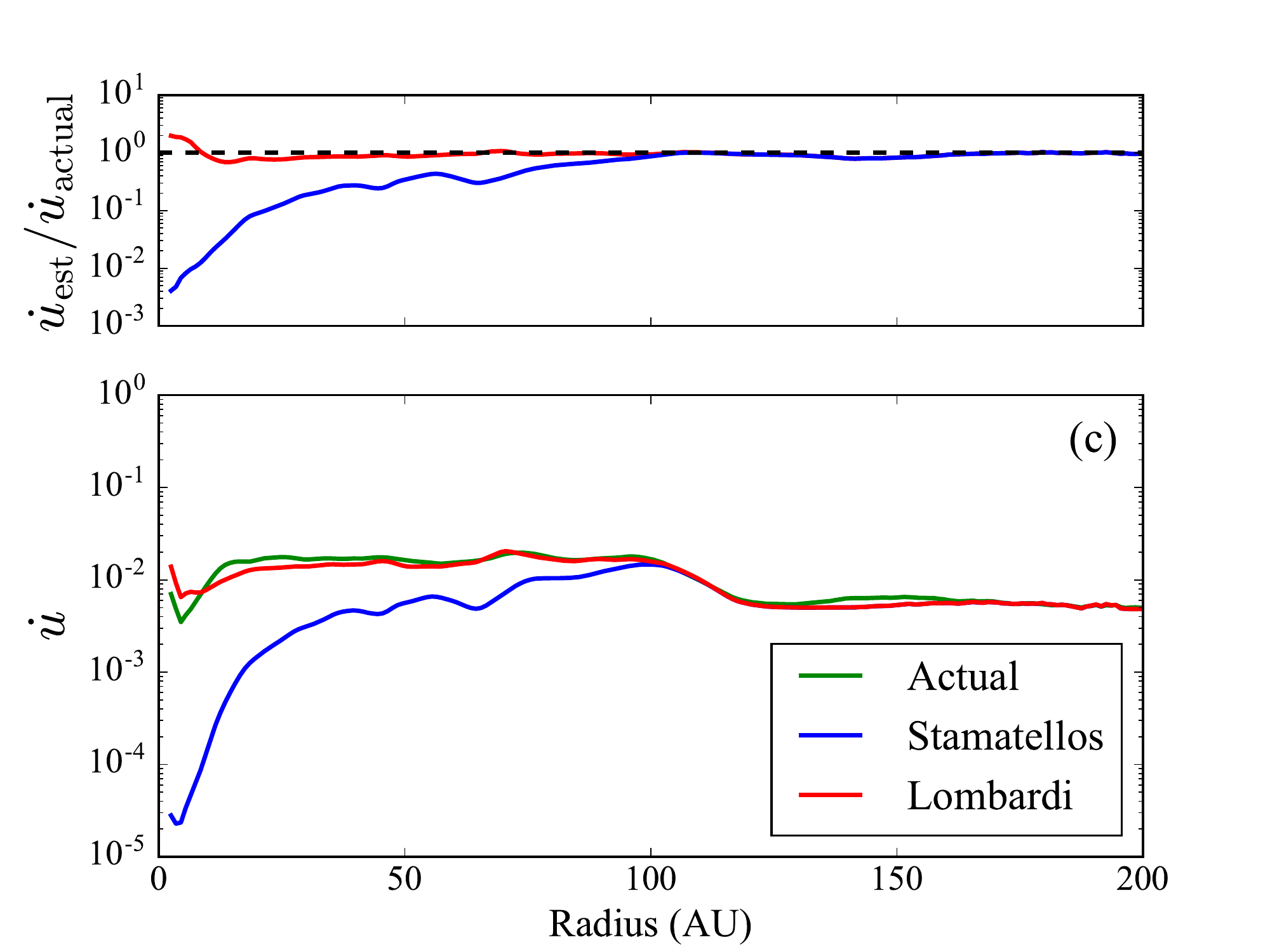}} \\
    \subfloat{\includegraphics[width =0.465\textwidth, trim = 0cm 0cm 0cm 0cm,
    clip=true]{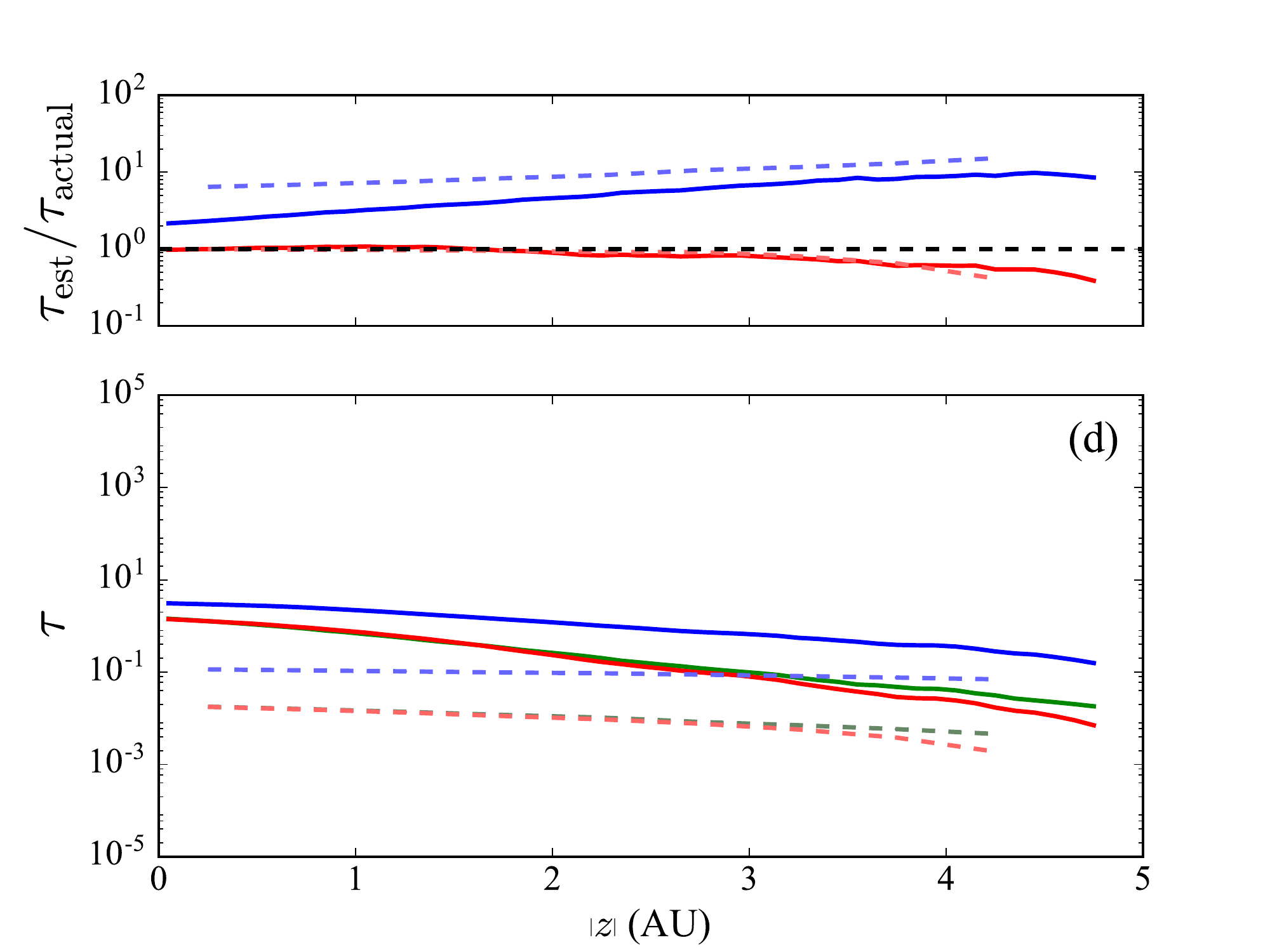}}
    \subfloat{\includegraphics[width =0.465\textwidth, trim = 0cm 0cm 0cm 0cm,
    clip=true]{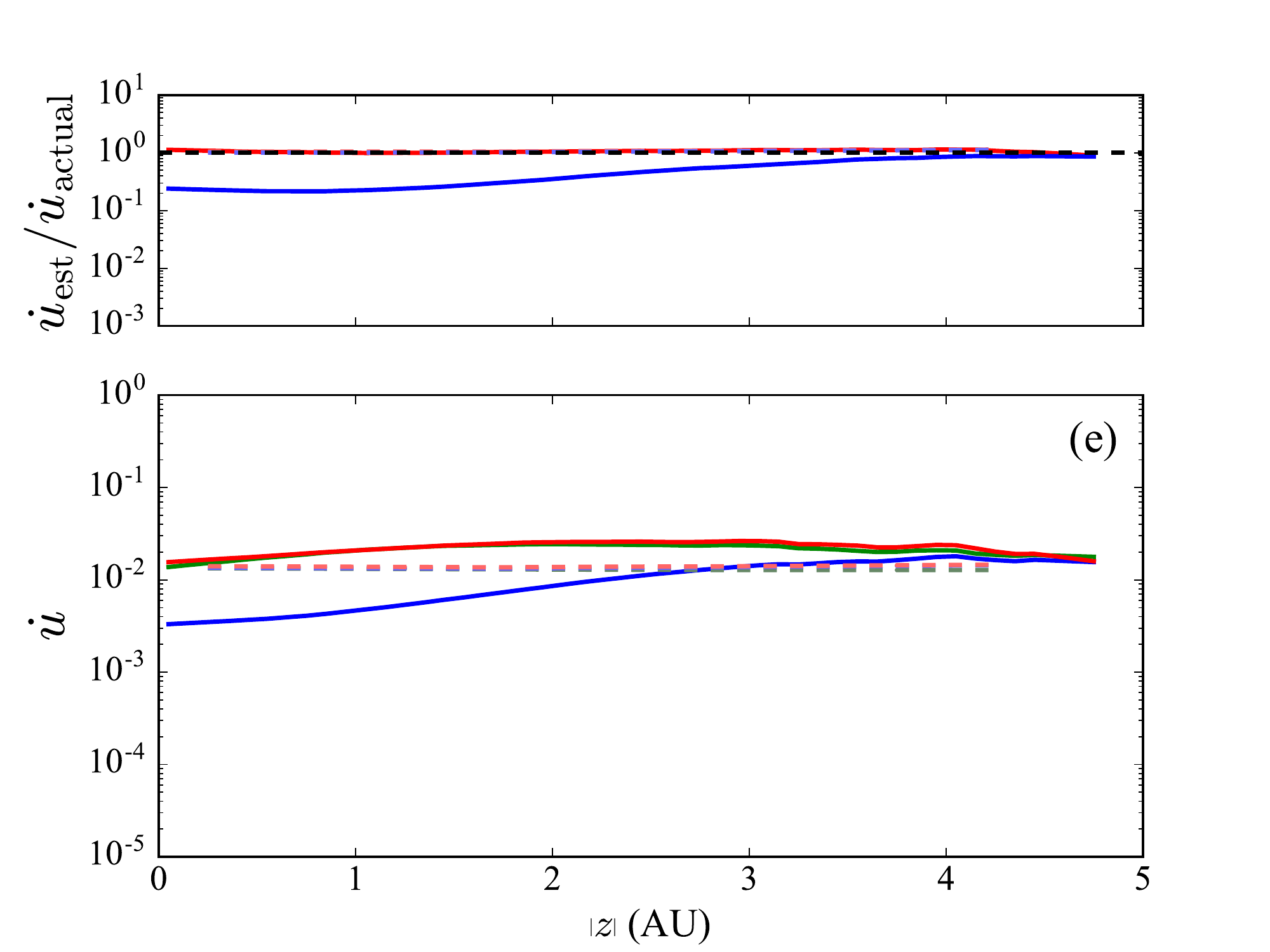}}
    \caption
    {
      A high-mass disc which has evolved to form spiral arms. Panel (a): a
      column density snapshot. White circles represent cylindrical regions where
      we perform an analysis perpendicular to the disc midplane. Panels (b) and
      (c): comparisons of azimuthally-averaged optical depth and cooling rate at
      the disc midplane. Panels (d) and (e): optical depth and cooling rate
      comparisons perpendicular to the disc midplane inside (solid lines), and
      outside (dashed lines) of a spiral arm. The upper plots in panels (b-e)
      show the ratio between estimated and actual values. The black dashed lines
      represent equality. The optical depth and cooling rate at the disc
      midplane are well estimated by the Lombardi method at all disc radii, but
      are over- and underestimated by the Stamatellos method, respectively.
      Vertically to the disc midplane, the same result is observed within a
      spiral arm. However, outside of the spiral arms, where the disc is
      optically thin, both methods yield a good estimate for the cooling rate.
    }
    \label{fig:gi_comparison}
    \end{centering}
\end{figure*}

After some time, the disc becomes unstable and spiral arms begin to form. This
is shown in Figure \ref{fig:gi_comparison}a. The optical depth and cooling rate
at the disc midplane are well described by the Lombardi method, but are over-
and underestimated, respectively, by the Stamatellos method. The cooling rate
estimated by the Stamatellos method is in agreement with the actual value
when the disc is optically thin (Figure \ref{fig:gi_comparison}b). We consider
two cylindrical regions with base radius of 5~AU wherein we perform vertical
analyses: one cylinder is inside a spiral arm and and the other outside (see
marked regions in Figure \ref{fig:gi_comparison}a). Outside the spiral arm, the
disc is optically thin and the cooling rate is estimated well by both methods
(Figure \ref{fig:gi_comparison}e, dashed lines). However, inside the spiral arm
where the disc is optically thick, the Stamatellos method overestimates
the optical depth and therefore the cooling rate. The Lombardi method
provides more accurate values for both quantities (Figure
\ref{fig:gi_comparison}e, solid lines).

% FIGURE : GI SNAPSHOT COMPARISON ----------------------------------------------

% ------------------------------------------------------------------------------
% DISC WITH CLUMPS
% ------------------------------------------------------------------------------
\subsection{High-mass disc with clumps}
\label{sub:disc_with_clumps}

\begin{figure*}
    \begin{centering}
    \subfloat{\includegraphics[width =0.465\textwidth, trim = 0cm 0cm 0cm 0cm,
    clip=true]{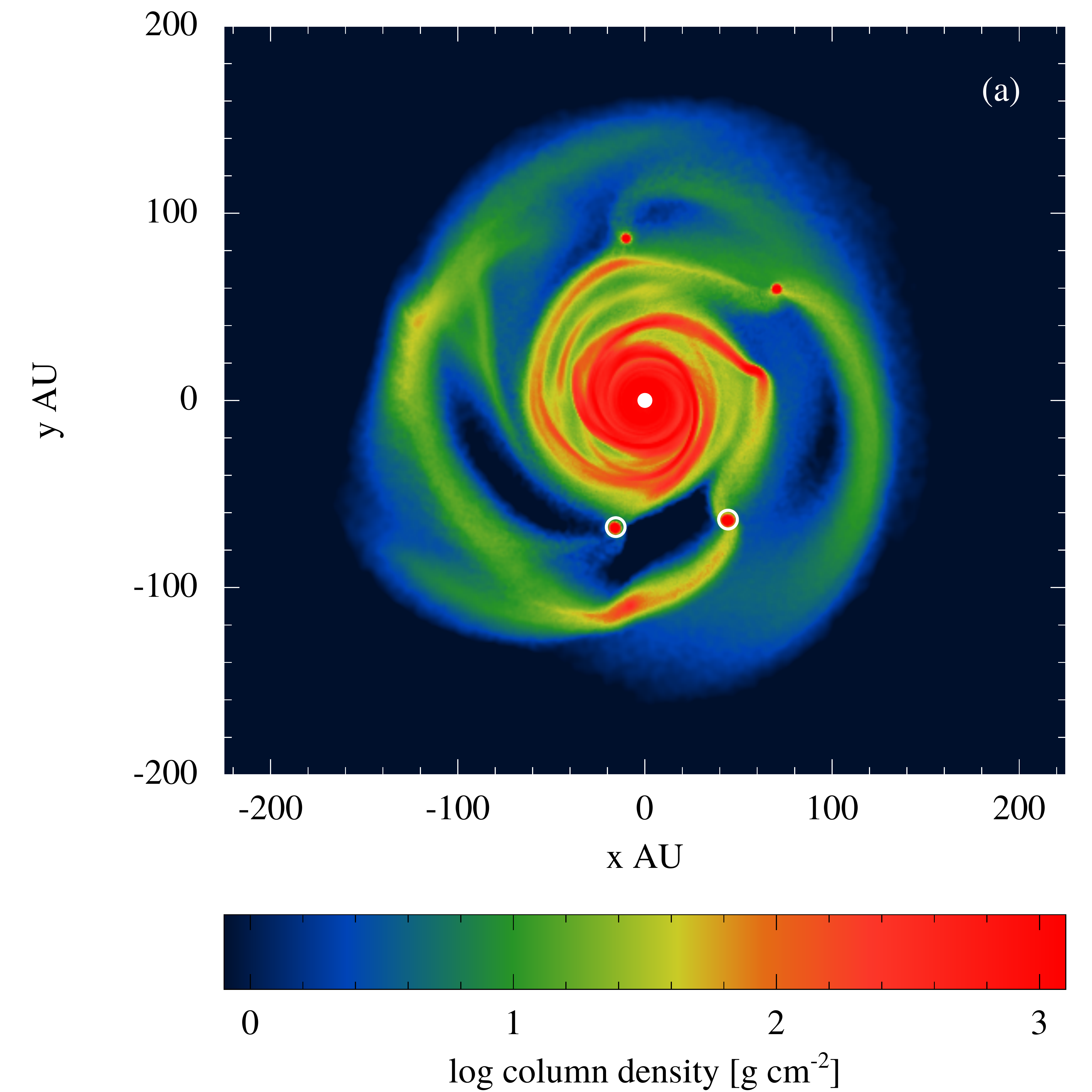}} \\
    \subfloat{\includegraphics[width =0.465\textwidth, trim = 0cm 0cm 0cm 0cm,
    clip=true]{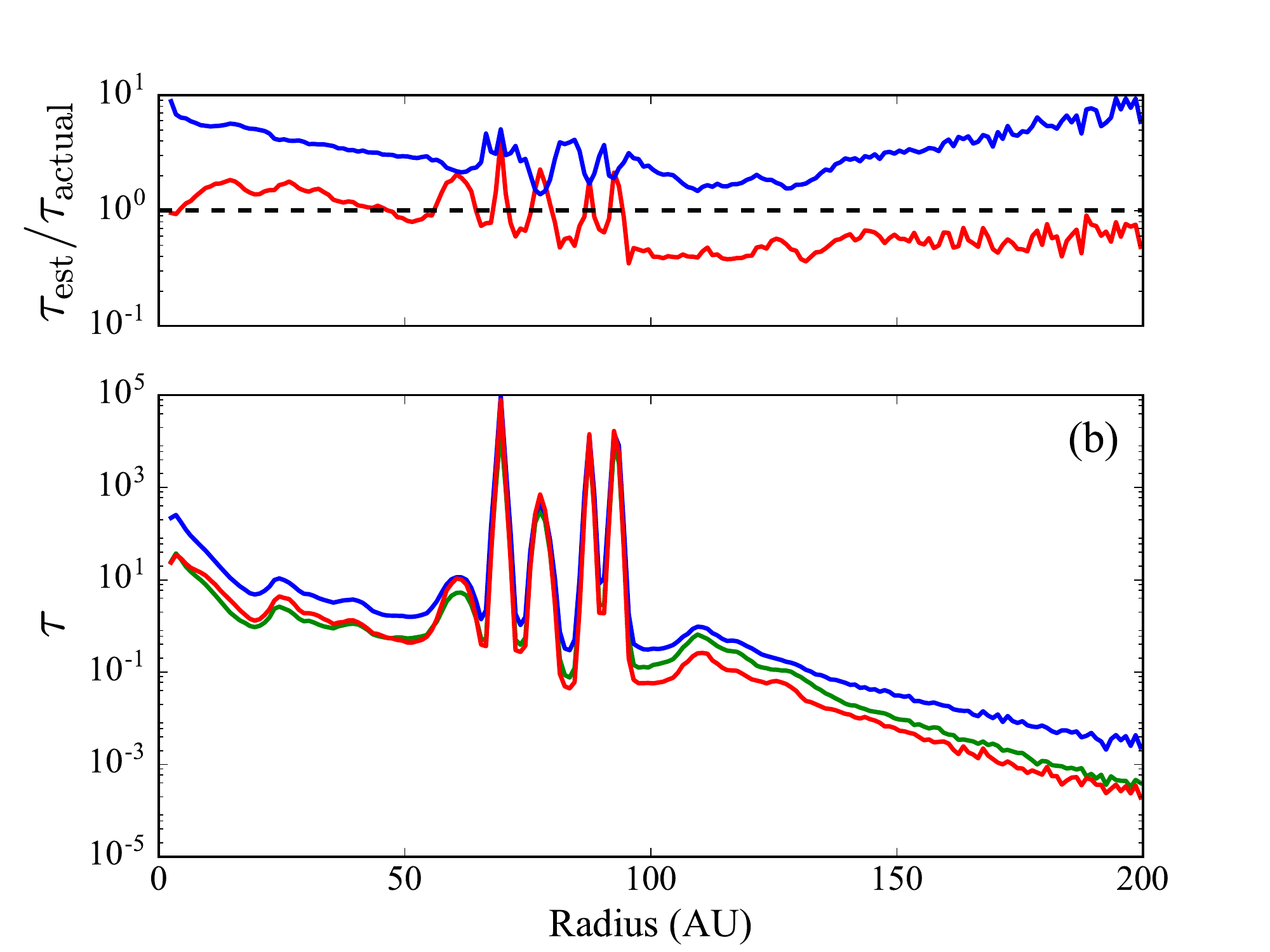}}
    \subfloat{\includegraphics[width =0.465\textwidth, trim = 0cm 0cm 0cm 0cm,
    clip=true]{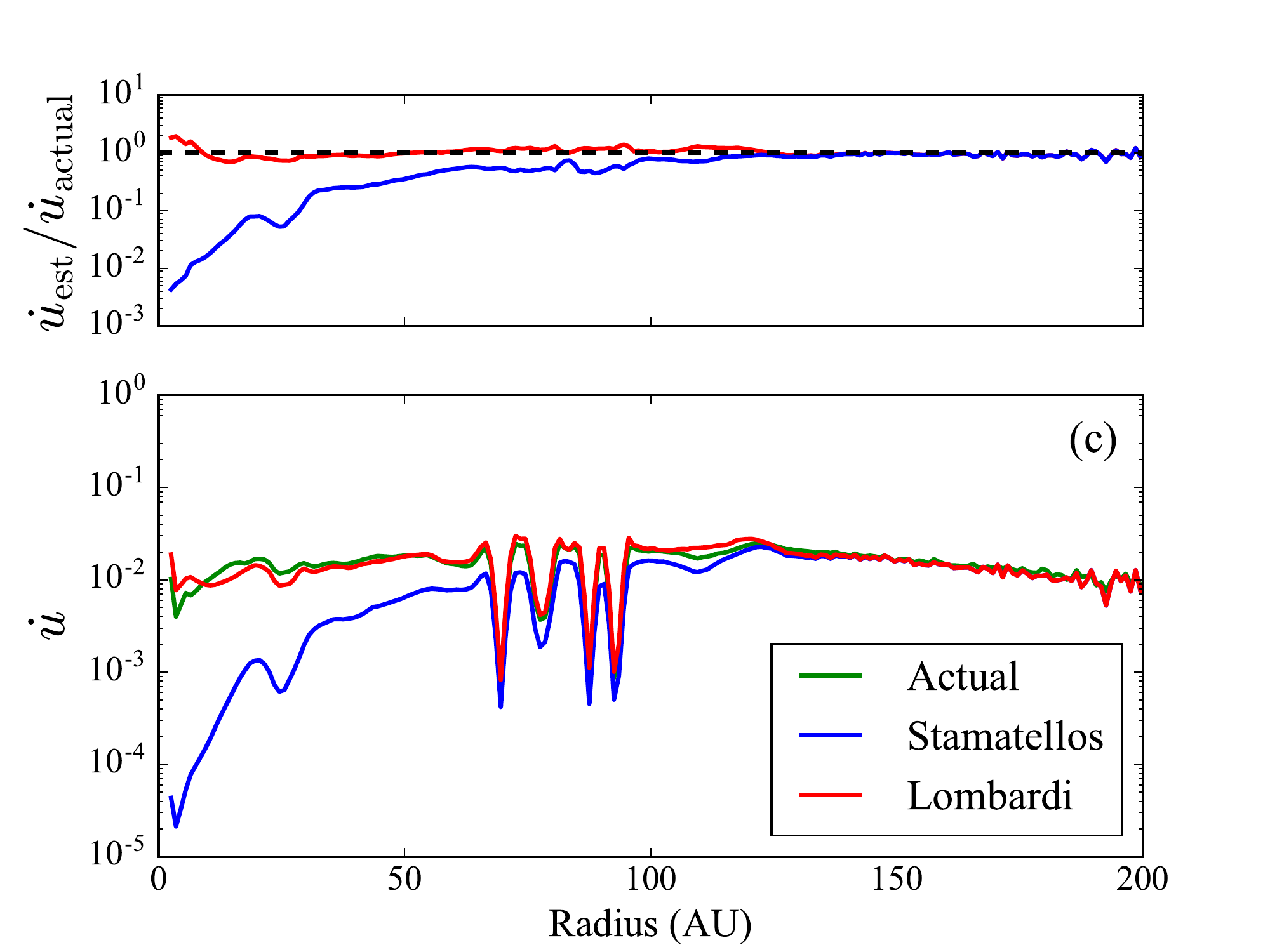}} \\
    \subfloat{\includegraphics[width =0.465\textwidth, trim = 0cm 0cm 0cm 0cm,
    clip=true]{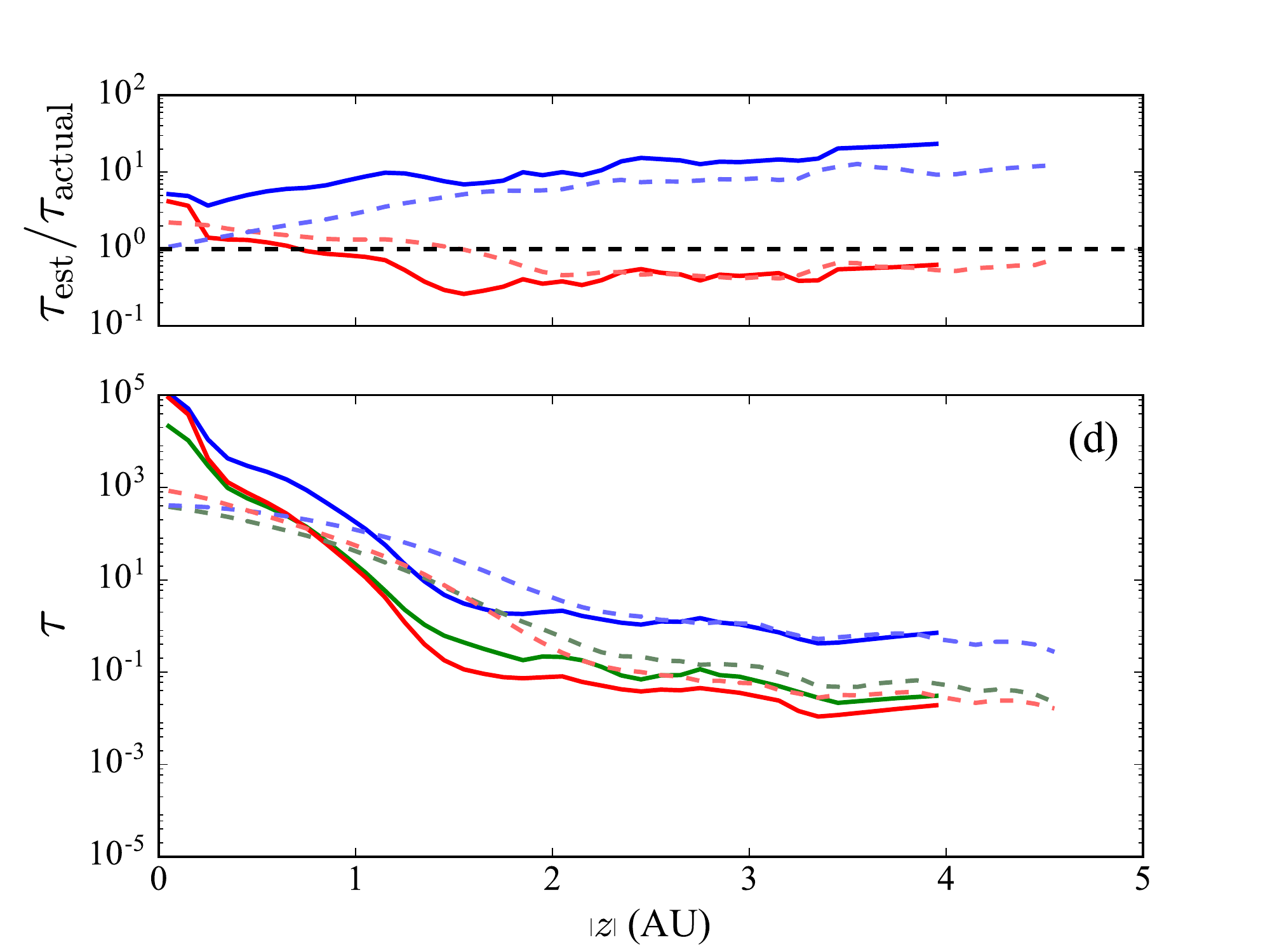}}
    \subfloat{\includegraphics[width =0.465\textwidth, trim = 0cm 0cm 0cm 0cm,
    clip=true]{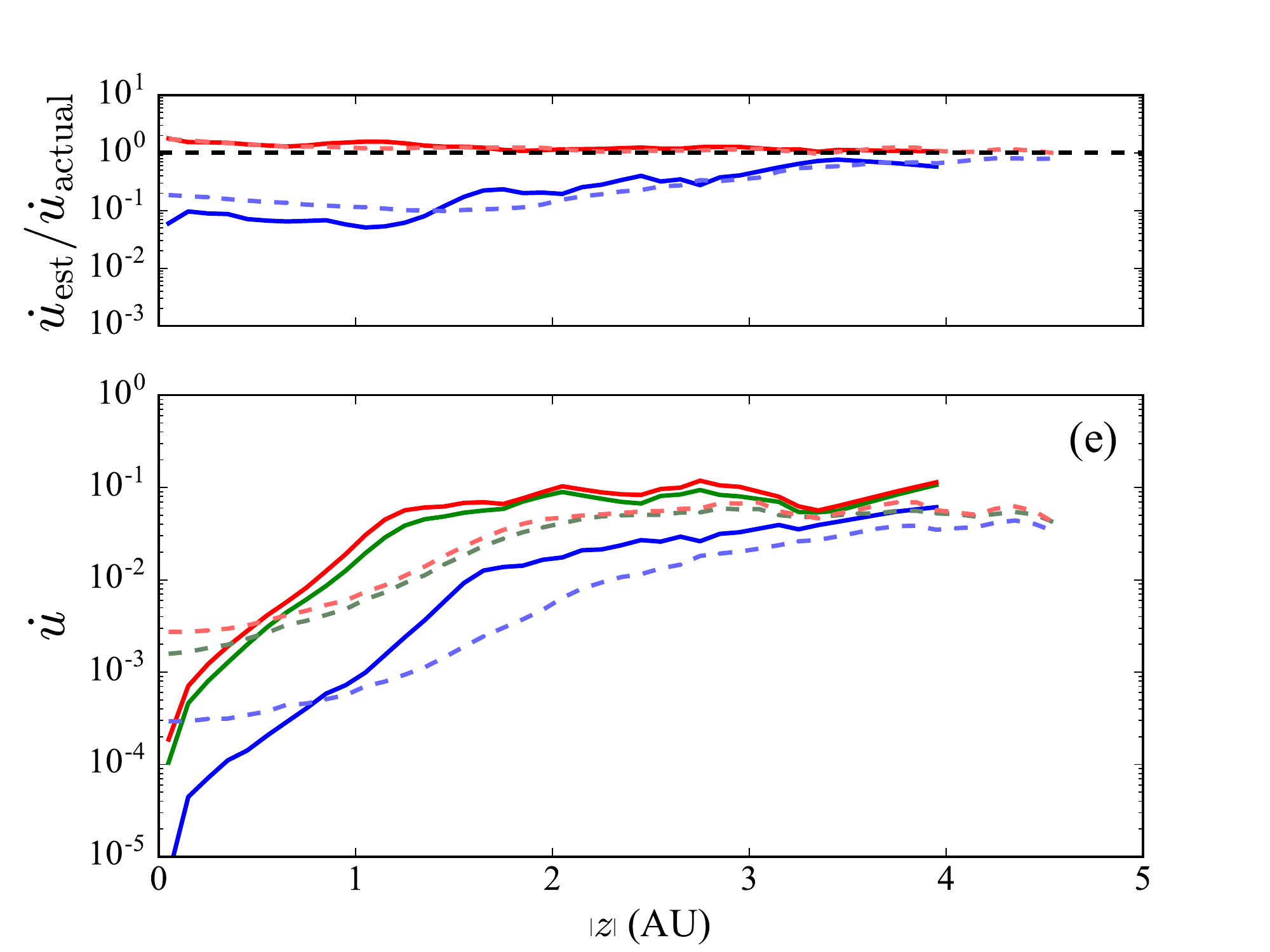}}
    \caption
    {
      A high-mass disc which has evolved to form dense clumps. Panel (a): a
      column density snapshot. White circles represent regions where vertical
      analyses are performed. Panels (b) and (c): comparisons of
      azimuthally-averaged optical depth and cooling rate at the disc midplane.
      Panels (d) and (e): optical depth and cooling rate comparisons
      perpendicular to the disc midplane for the densest clump (solid lines),
      and the least dense clump (dashed lines). The upper plots in panels (b-e)
      show the ratio between estimated and actual values. The black dashed lines
      represent equality. The optical depth is generally overestimated by the
      Stamatellos method. The Lombardi method gives a better estimate,
      even within the dense clump. The cooling rate is also estimated more
      accurately.
    }
    \label{fig:dense_comparison}
    \end{centering}
\end{figure*}

The disc eventually fragments and dense clumps form. The column density snapshot
in Figure \ref{fig:dense_comparison}a contains four clumps. The central density
of the densest clump is $\sim 10^{-6} \textup{ g cm}^{-3}$ and for the least
dense clump is $\sim 10^{-10} \textup{ g cm}^{-3}$. Figure
\ref{fig:dense_comparison}b shows that both the Stamatellos and Lombardi methods
give good estimates of the azimuthally-averaged optical depth at the disc
midplane, but it should be noted that an azimuthally-averaged analysis is not
ideal for describing this disc, as it is highly non-axisymmetric. Therefore we
focus on two of the clumps: the inner, densest clump, and the least dense clump.
We consider a cylinder with base radius of 5 AU centred on each of these clumps
and we perform a vertical analysis in the direction perpendicular to the disc
midplane. Figure \ref{fig:dense_comparison}d shows the optical depth comparison.
We find that for the least dense clump (dashed lines), the Stamatellos method is
accurate in the centre of the clump. The Lombardi method overestimates the
optical depth by a factor $\sim 2$. In the centre of the densest clump, both
methods are inaccurate, but only by a factor of a few. In general - for the disc
as a whole as well as the clumps - the Lombardi method estimates the cooling
rate well, whilst the Stamatellos method systematically underestimates the
cooling rate.

% FIGURE : DENSE SNAPSHOT COMPARISON -------------------------------------------

% FIGURE : LOWER-MASS PLANET GAP -----------------------------------------------
\begin{figure*}
    \begin{centering}
    \subfloat{\includegraphics[width =0.465\textwidth, trim = 0cm 0cm 0cm 0cm,
    clip=true]{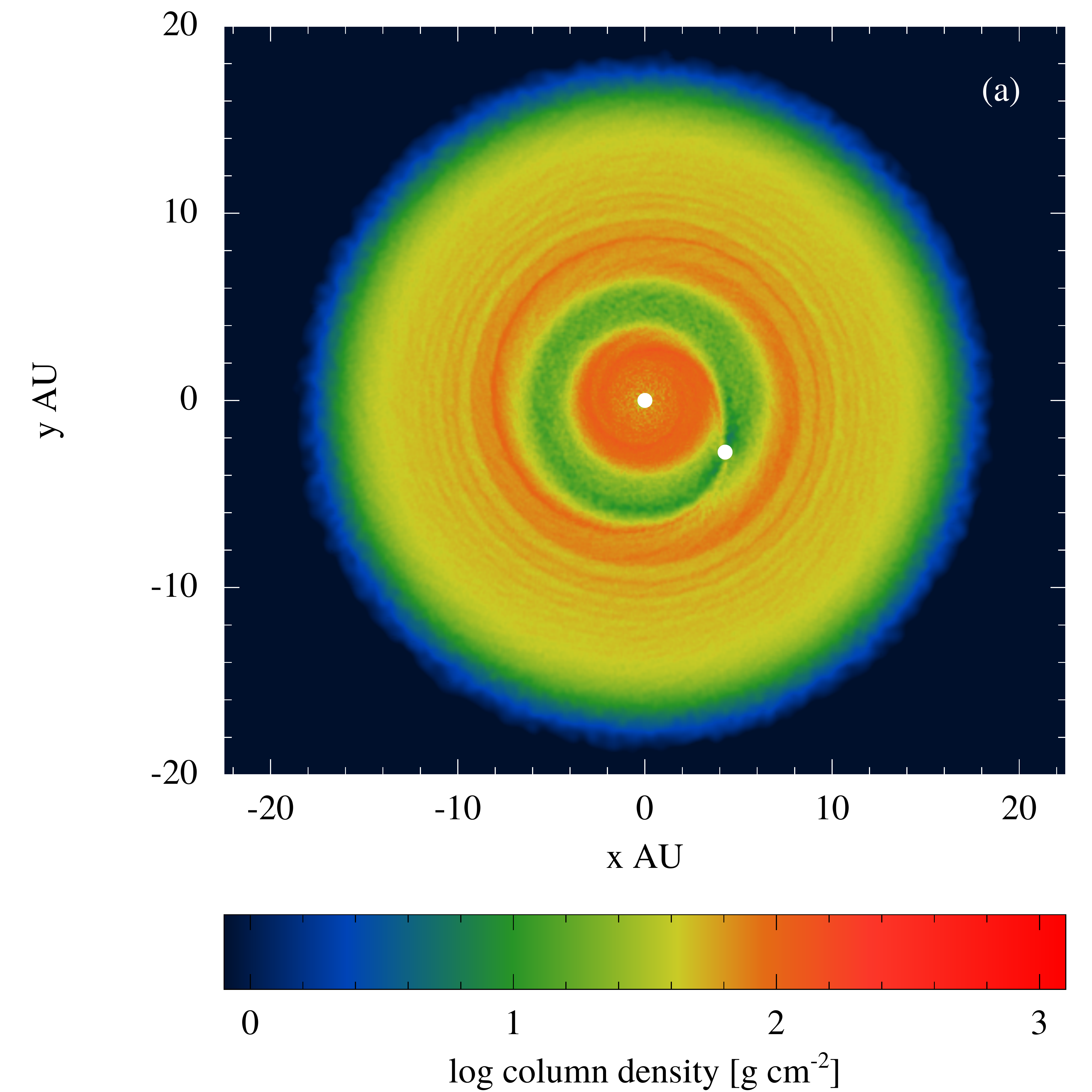}} \\
    \subfloat{\includegraphics[width =0.465\textwidth, trim = 0cm 0cm 0cm 0cm,
    clip=true]{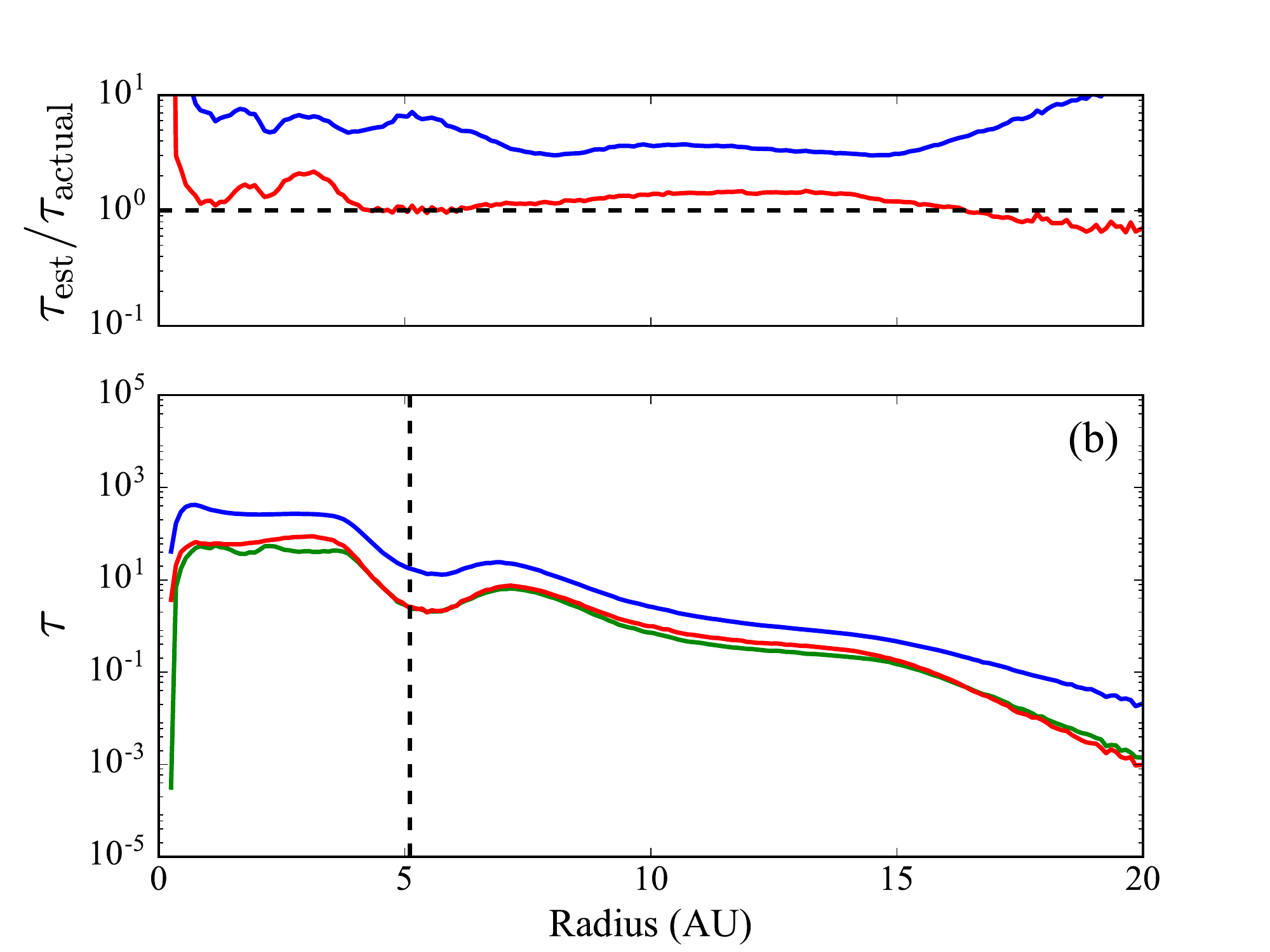}}
    \subfloat{\includegraphics[width =0.465\textwidth, trim = 0cm 0cm 0cm 0cm,
    clip=true]{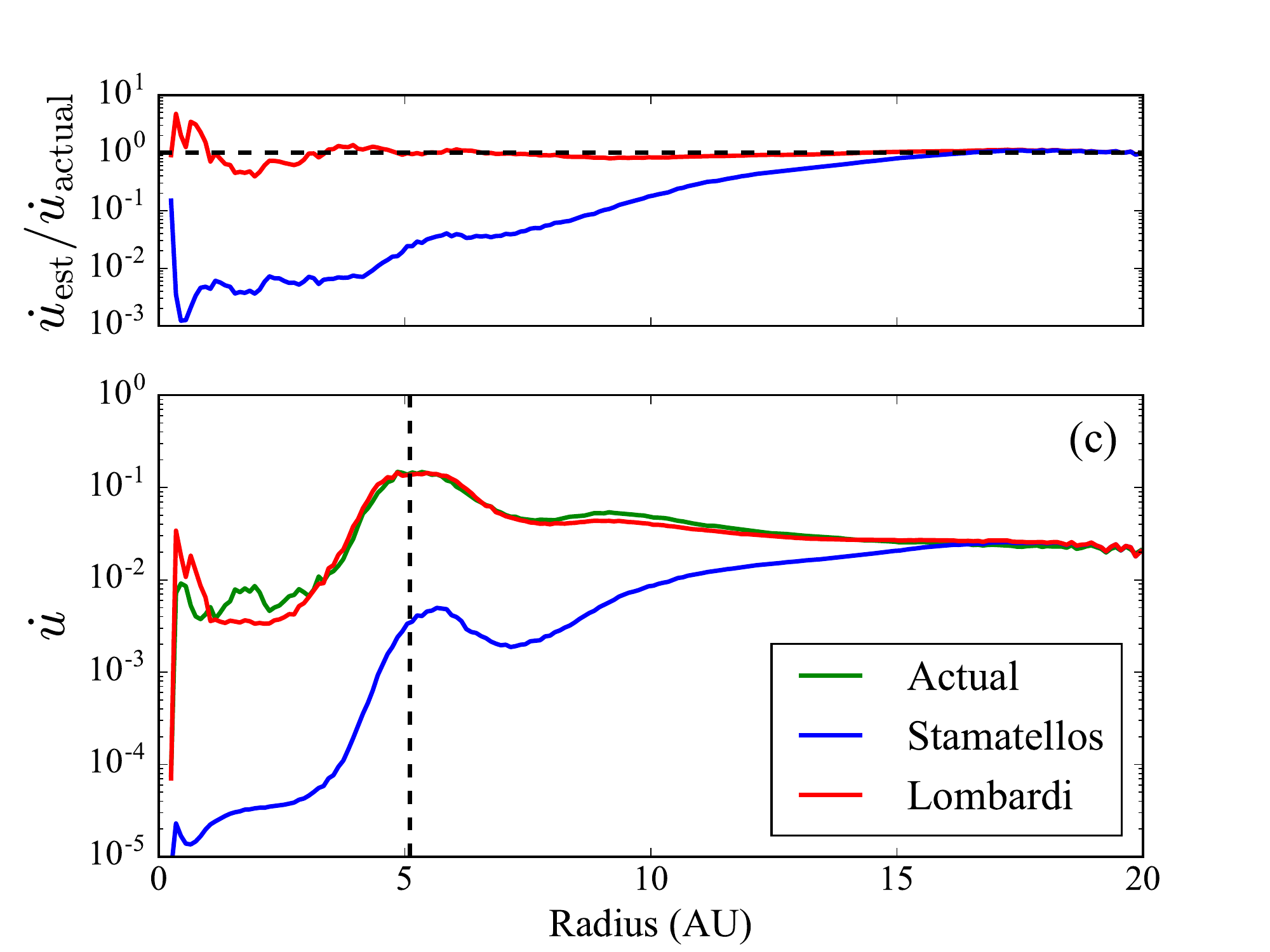}} \\
    \subfloat{\includegraphics[width =0.465\textwidth, trim = 0cm 0cm 0cm 0cm,
    clip=true]{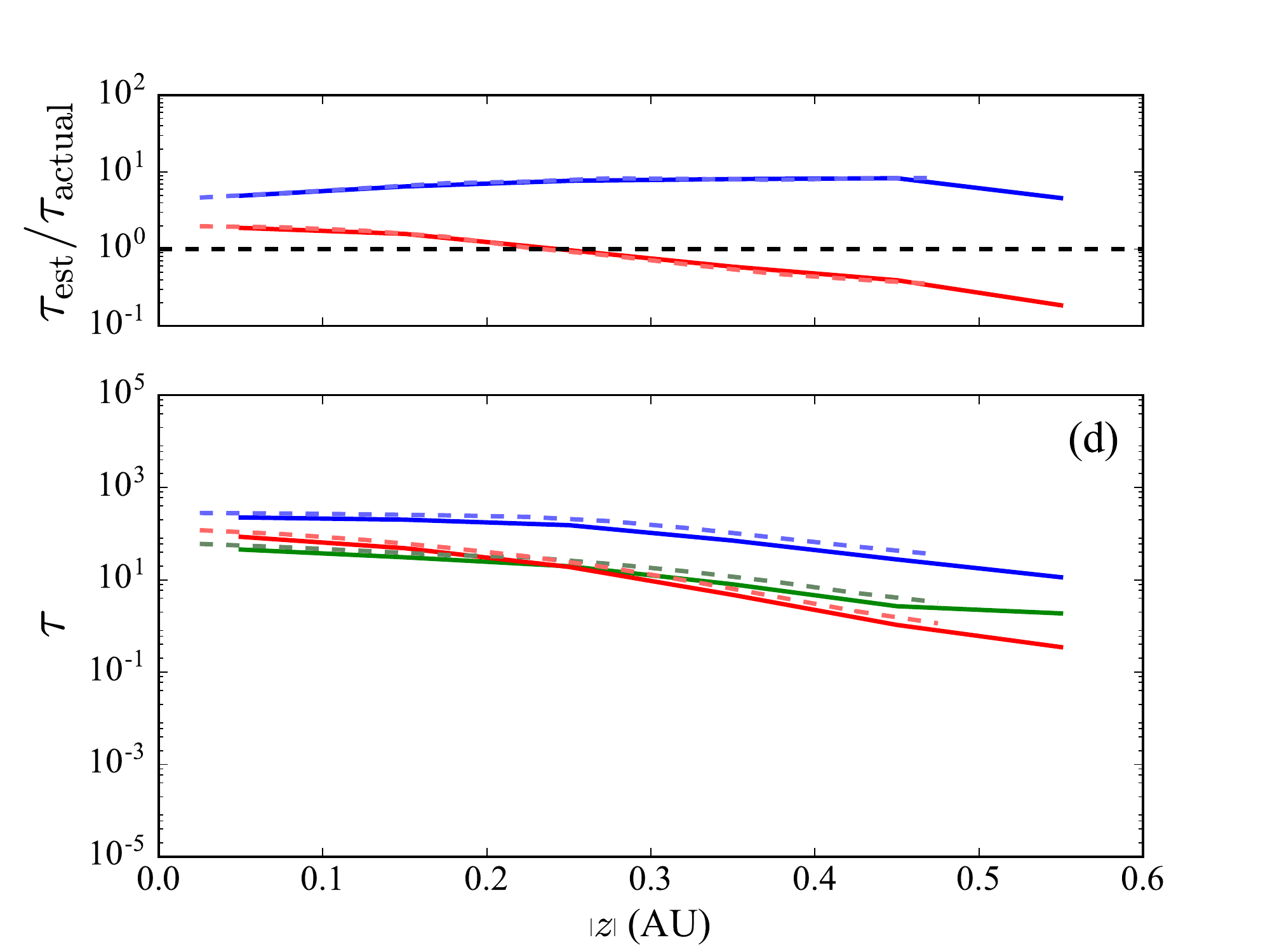}}
    \subfloat{\includegraphics[width =0.465\textwidth, trim = 0cm 0cm 0cm 0cm,
    clip=true]{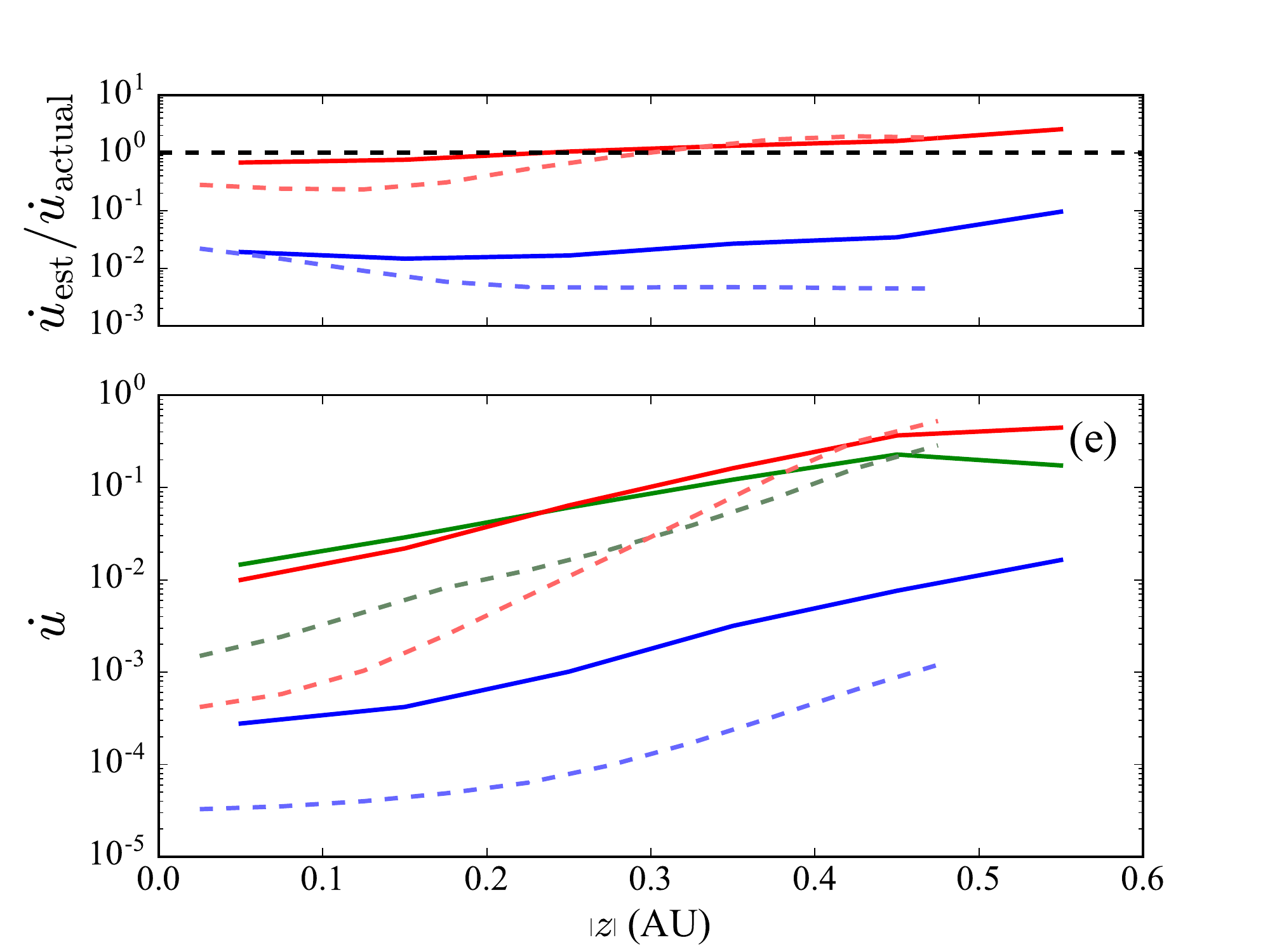}}
    \caption
    {
      A disc which has an embedded $1.4 \mjup$ planet at a radius of $5.1$~AU.
      Panel (a): a column density snapshot. Panels (b) and (c): comparisons of
      azimuthally-averaged optical depth and cooling rate at the disc midplane.
      The vertical black dashed lines in panels (b) and (c) represent the
      location of the planet. Panels (d) and (e): optical depth and cooling rate
      comparisons perpendicular to the disc midplane between radial annuli of
      $4-6$~AU (in the gap, solid lines), and $3-4$~AU (interior to the gap,
      dashed lines). Gas within $R_{\textsc{hill}} = 0.6$~AU of the planet is
      excluded when analysing the gap. The upper plots in panels (b-e) show the
      ratio between estimated and actual values. The black dashed lines
      represent equality. The Stamatellos method overestimates the optical depth
      by a factor of 3 or more throughout the disc. The Lombardi method
      estimates the optical depth within a factor of 2, and it also gives an
      accurate estimate of the cooling rate, both inside and outside the
      planet-induced gap.
    }
    \label{fig:gapl_comparison}
    \end{centering}
\end{figure*}

% ==============================================================================
% PROTOSTELLAR DISCS WITH EMBEDDED PLANETS
% ==============================================================================
\section{Protostellar discs with embedded planets}
\label{sec:protostellar_discs_with_embedded_planets}

The gravitational interaction between a planet and the surrounding disc may
result in the formation of planet-induced gaps \citep[e.g.][]{Goldreich:1980a,
Lin:1993a, Bryden:1999a, Kley:2012b}. Such structures may provide indirect
evidence for the presence of planets in discs. The \cite{Crida:2006a}
semi-analytical criterion for gap formation involves the balance between the
tidal torque which opens the gap and the viscous torque which closes the gap. It
has been shown that planets with masses down to $10 \mearth$ can open gaps
\citep{Duffell:2012a}. However, for migrating planets, a gap must form on a
rapid enough timescale. \cite{Malik:2015a} argue that a gap can only form
provided the gap opening time is longer than the migration timescale of the
planet. The accurate treatment of the radiative transfer
in such planet-disc systems is important and may play a significant role when
determining the rate and the direction (i.e. inwards or outwards) of migration,
and the final mass of the planet
\citep{Stamatellos:2015a,Benitez-Llambay:2015a,Stamatellos:2018a}.

Here we examine two cases of protostellar discs with embedded planets: one with
an embedded $1.4 \mjup$ planet (\textsection
\ref{sub:disc_with_a_low_mass_planet}) and one with an embedded higher-mass, $11
\mjup$, planet (\textsection \ref{sub:disc_with_a_high_mass_planet}). We compare
the estimated optical depth and cooling rate obtained via the
\cite{Stamatellos:2007b} and \cite{Lombardi:2015a} radiative transfer methods.

% ------------------------------------------------------------------------------
% DISC WITH A LOWER-MASS PLANET
% ------------------------------------------------------------------------------
\subsection{Disc with an embedded $1.4 \mjup$ planet}
\label{sub:disc_with_a_low_mass_planet}

We consider a disc with an initial mass $0.005 \msun$ surrounding a $1 \msun$
protostar. A $1\mjup$ mass planet is embedded within the disc at a radius of
5.2~AU. The initial disc extends out to 15.6~AU, with a surface density profile
$\Sigma(R)\propto R^{-1/2}$ \citep[e.g.][]{Bate:2003a}, temperature profile
$T(R) \propto R^{-3/4}$, and is represented by $10^6$ SPH particles. The
temperature at 1 AU from the central star is $T_{0} = 250$~K. The planet
migrates slightly inwards (0.1 AU) and increases in mass by accreting
gas from the disc. At the snapshot presented here (Figure
\ref{fig:gapl_comparison}a) the planet is at 5.1~AU and has carved out a gap
between 4 and 6 AU. Its mass has increased to $1.4\mjup$.

The density of the disc is high and as such, the disc is optically thick (Figure
\ref{fig:gapl_comparison}b). The Stamatellos method overestimates the optical
depth at the disc midplane throughout the disc by a factor of a few, whilst the
Lombardi method provides a better estimate (accurate within a factor of $\sim$
2). This is reflected in the estimated cooling rates (Figure
\ref{fig:gapl_comparison}c). 

Vertical profiles are shown for radial annuli at the planet gap ($4-6$ AU;
Figure \ref{fig:gapl_comparison}d, e - solid lines) as well as on a region
interior to the gap ($3-4$ AU, Figure \ref{fig:gapl_comparison}d, e - dashed
lines). We exclude gas within the Hill radius ($R_{\textsc{hill}} = 0.6$~AU) of
the planet when analysing the gap region. Both of these regions are optically
thick. Again, the Lombardi method provides a better estimate for the optical
depth and cooling rate. 

In the gap region, which is important for the evolution of the planet, the
Lombardi method is very accurate, whereas the Stamatellos method overestimates
the optical depth, and therefore underestimates the cooling rate. 

% ------------------------------------------------------------------------------
% DISC WITH A HIGH-MASS PLANET
% ------------------------------------------------------------------------------
\subsection{Disc with an embedded $11 \mjup$ planet}
\label{sub:disc_with_a_high_mass_planet}

\begin{figure*}
    \begin{centering}
    \subfloat{\includegraphics[width =0.465\textwidth, trim = 0cm 0cm 0cm 0cm,
    clip=true]{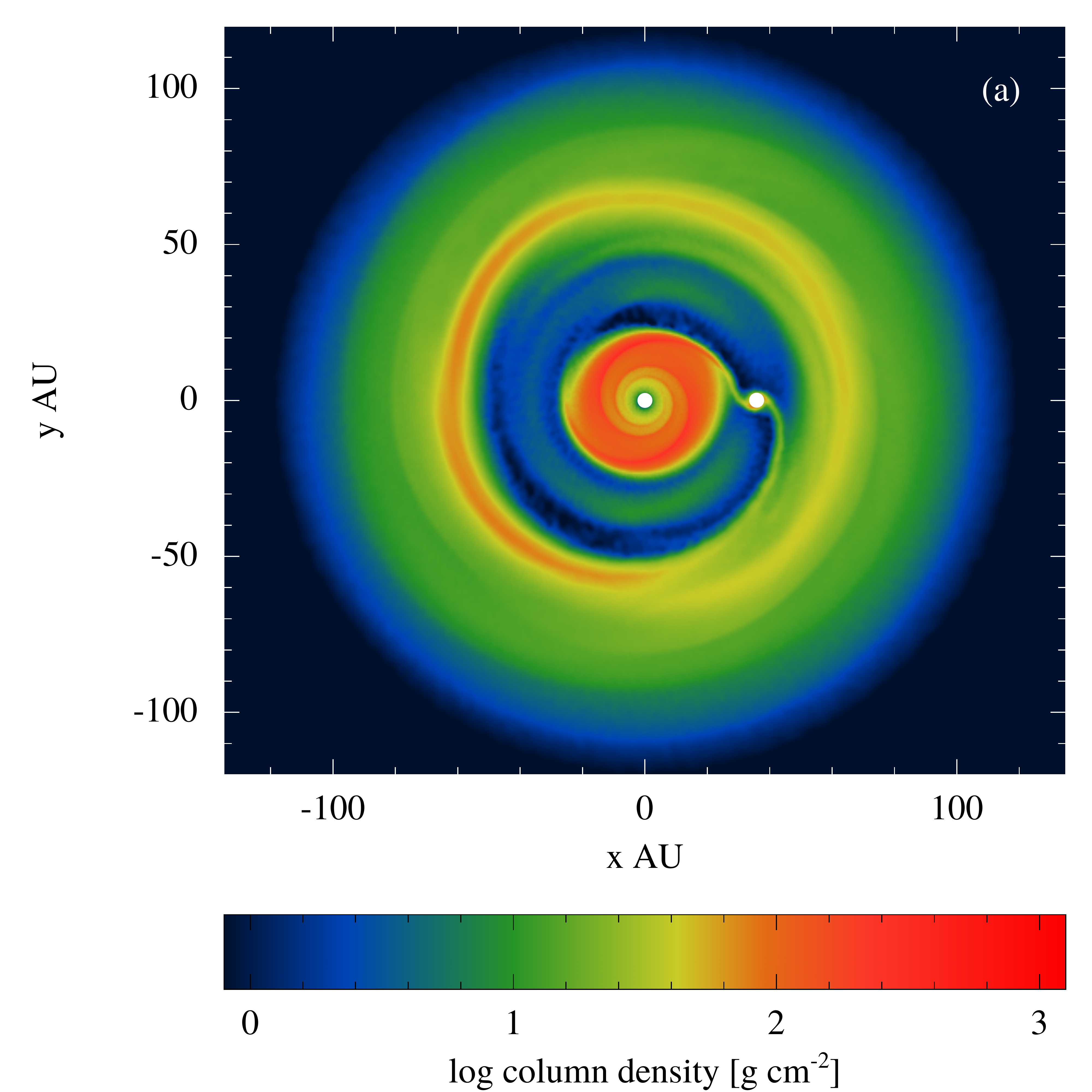}} \\
    \subfloat{\includegraphics[width =0.465\textwidth, trim = 0cm 0cm 0cm 0cm,
    clip=true]{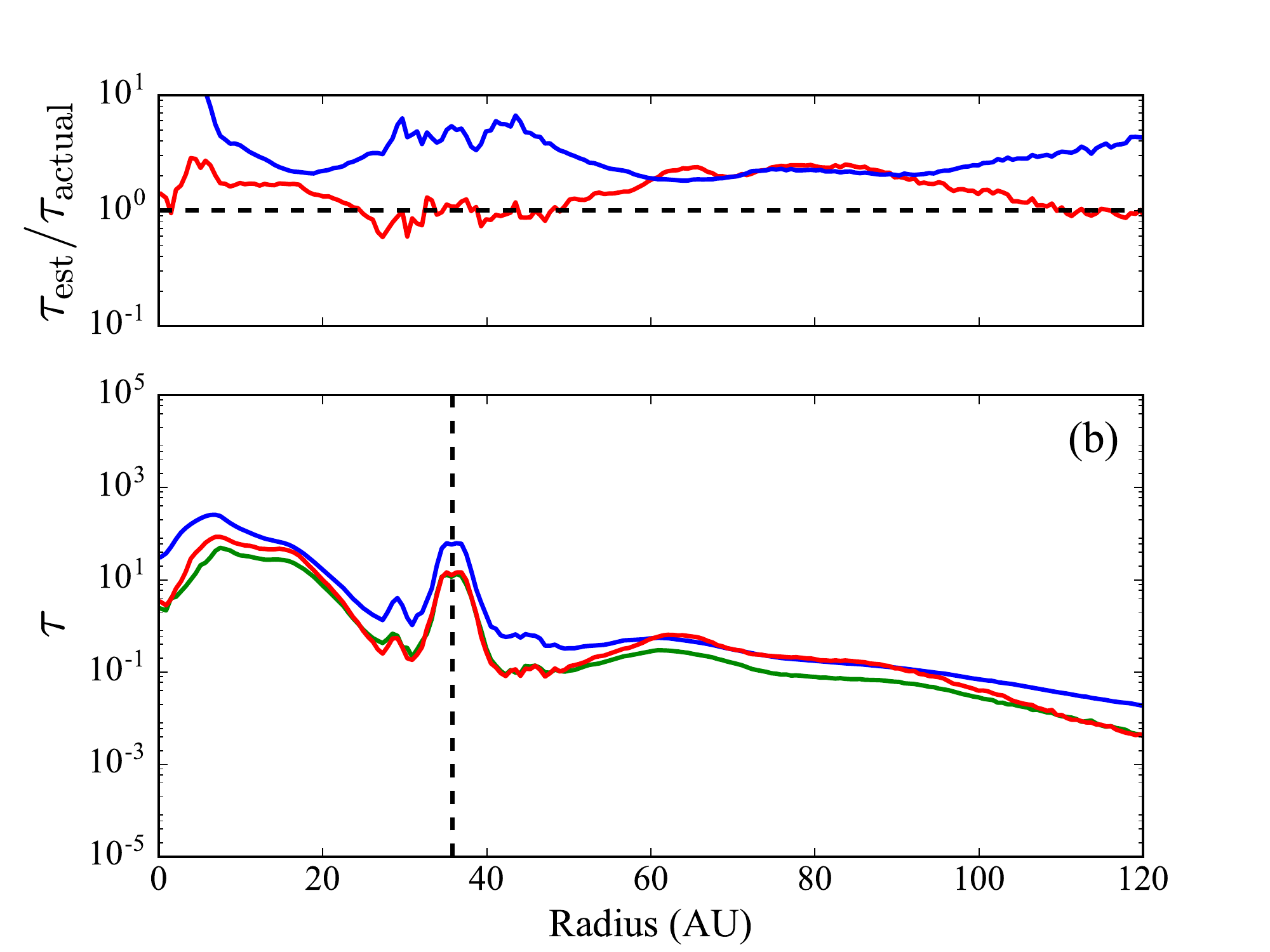}}
    \subfloat{\includegraphics[width =0.465\textwidth, trim = 0cm 0cm 0cm 0cm,
    clip=true]{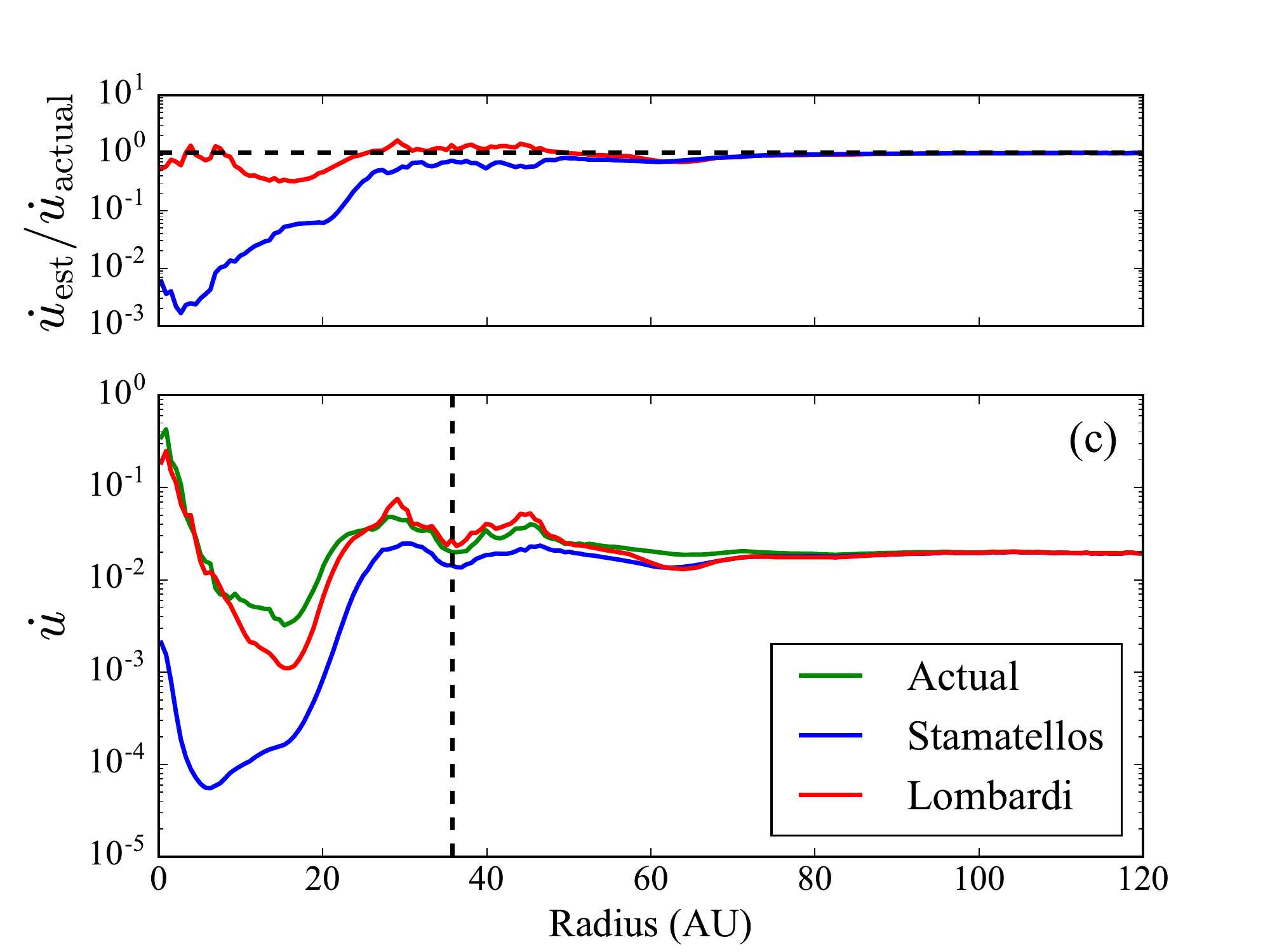}} \\
    \subfloat{\includegraphics[width =0.465\textwidth, trim = 0cm 0cm 0cm 0cm,
    clip=true]{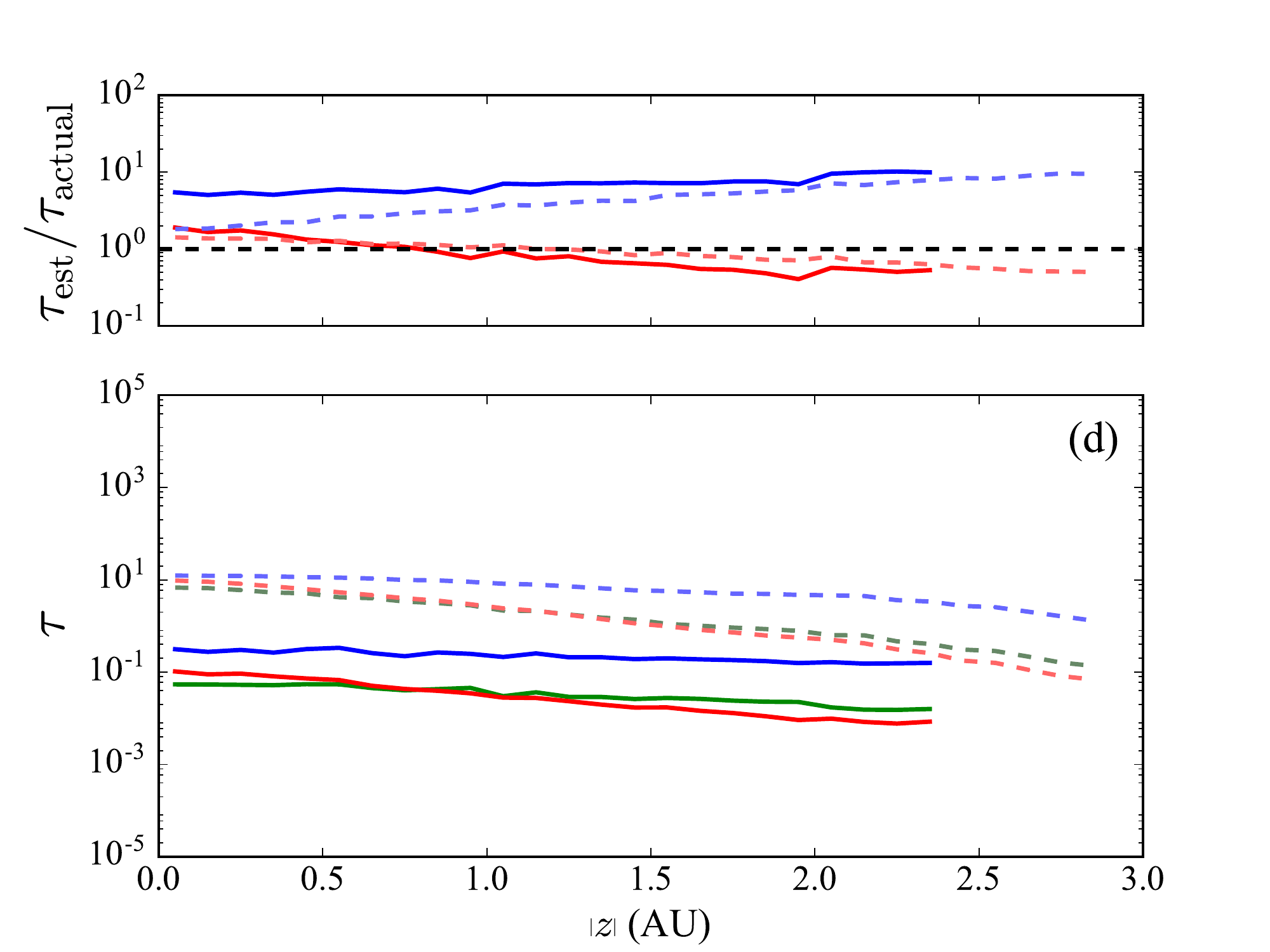}}
    \subfloat{\includegraphics[width =0.465\textwidth, trim = 0cm 0cm 0cm 0cm,
    clip=true]{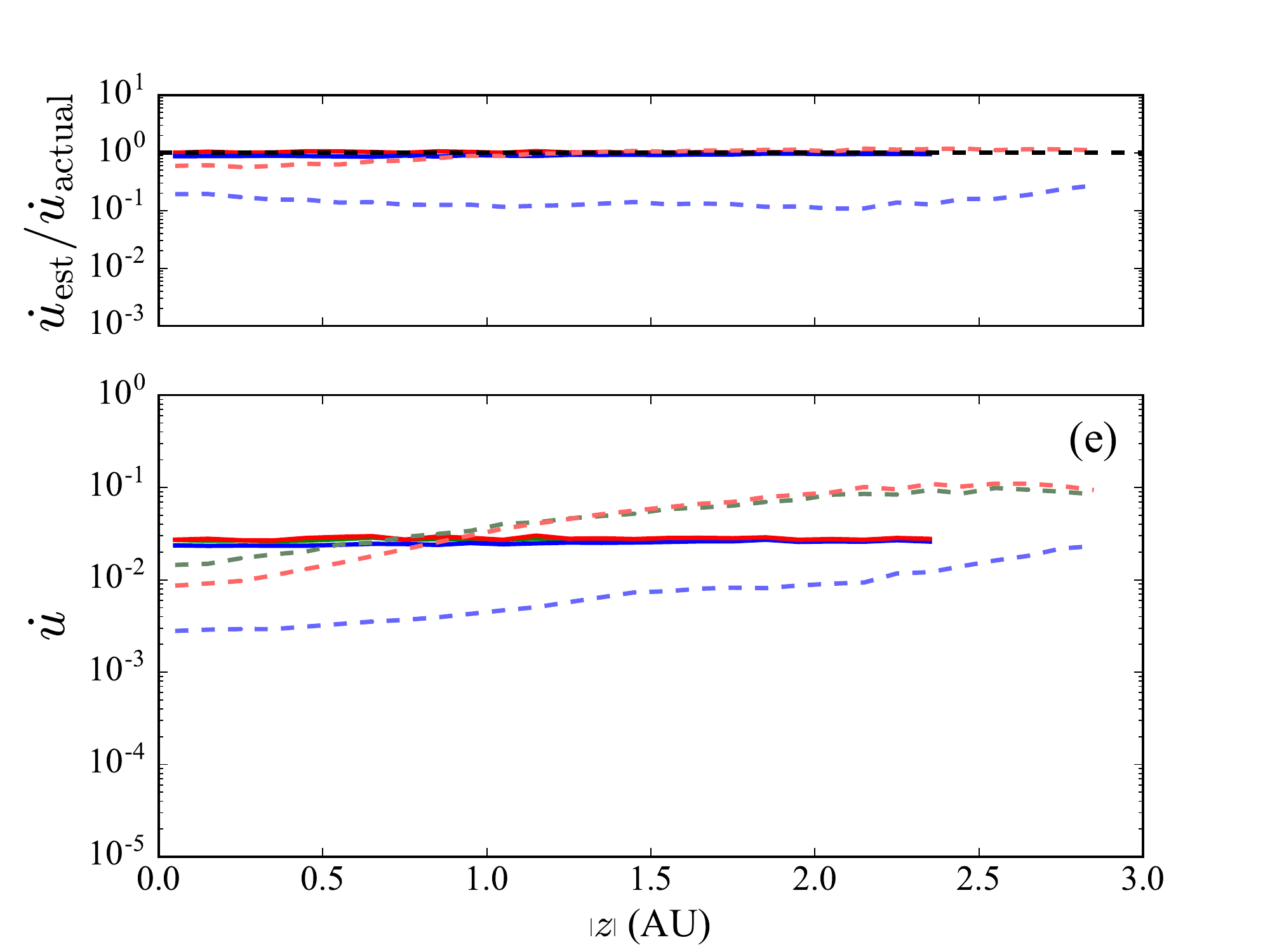}}
    \caption
    {
      A disc which has an embedded $11 \mjup$ planet at a radius of $36$ AU.
      Panel (a): a column density snapshot. Panels (b) and (c): comparisons of
      azimuthally-averaged optical depth and cooling rate at the disc midplane.
      The vertical black dashed lines in panels (b) and (c) represent the
      location of the planet. Panels (d) and (e): optical depth and cooling rate
      comparisons perpendicular to the disc midplane between radial annuli of
      $33-37$~AU (inside the gap, solid lines), and $23-27$~AU (outside the gap,
      dashed lines). Gas within $R_{\textsc{hill}} = 8.0$~AU of the planet is
      excluded when analysing the gap. The upper plots in panels (b-e) show the
      ratio between estimated and actual values. The black dashed lines
      represent equality. Both methods overestimate the optical depth in the
      outer disc by a factor of 2 or 3. However, the Lombardi method estimates
      both the optical depth and the cooling within the gap more accurately than
      the Stamatellos method. Outside and within the gap, the Lombardi method
      gives a good estimate for both quantities from the disc midplane to the
      disc surface. The Stamatellos method estimates the cooling rate well
      within the gap as this region is optically thin.
    }
    \label{fig:gapm_comparison}
    \end{centering}
\end{figure*}

We simulate a system comprising a star which has an initial mass $1 \msun$, that
is attended by a protostellar disc with mass $0.1 \msun$ and initial radius
$100$~AU. The disc is modelled by $10^6$ SPH particles, and has initial surface
density and temperature profiles $\Sigma(R)\propto R^{-1}$ and $T(R) \propto
R^{-3/4}$, respectively \citep{Stamatellos:2015a}. The temperature at 1 AU from
the central star is $T_{0} = 250$~K. A planet with an initial mass $1 \mjup$ is
embedded in the disc at radius of $50$~AU. At the snapshot we present (Figure
\ref{fig:gapm_comparison}a) the disc mass has dropped to $0.08 \msun$ and the
planet mass has increased to $11 \mjup$. The planet has migrated inwards and is
located at a radial distance of $36$~AU. It has carved a gap between $\sim 30$
and $\sim 40$~AU. 

Figure \ref{fig:gapm_comparison}b shows that the Lombardi method estimates the
optical depth at the midplane of the disc well within the gap, but overestimates
it by a factor of a few outside of the gap. The Stamatellos method overestimates
the optical depth at all radii: by a factor of $\sim 2$ outside of the gap and
$\sim 10$ within the gap. 

We consider two radial annuli where we perform vertical analyses. One includes
the gap (between 33 and 37 AU, Figure \ref{fig:gapm_comparison}d, e - solid
lines), the other a region interior to the gap (between 23 and 27 AU, Figure
\ref{fig:gapm_comparison}d, e - dashed lines). The disc is optically thin within
the gap. Thus the cooling rate is well estimated by both methods. We exclude gas
within the Hill radius of the planet ($R_{\textsc{hill}} = 8.0$~AU) when
analysing the gap. The region interior to the gap is optically thick. The
cooling rate is well estimated at all $z$ by the Lombardi method, but the
Stamatellos method underestimates the cooling rate by up to a factor of 10.

% FIGURE : MASSIVE GAP SNAPSHOT COMPARISON -------------------------------------

% ==============================================================================
% BETA-COOLING COMPARISON
% ==============================================================================
\section{Testing the $\beta$-cooling approximation}
\label{sec:comparison_with_the_beta_cooling_approximation}

The $\beta$-cooling approximation \citep[e.g.][]{Gammie:2001a, Rice:2003a} is a
computationally inexpensive technique used when simulating accretion discs. This
method assumes that the cooling rate at a given radius $R$ within the disc,
is inversely proportional to cooling time such that
\begin{equation}
  \dot{u} = \frac{u}{t_{\textup{cool}}},
\label{eqn:beta_cooling_approx}
\end{equation}
where the cooling time is
\begin{equation}
  t_{\textup{cool}} = \beta \Omega^{-1}.
\label{eqn:cooling_time}
\end{equation}
$\Omega$ is the Keplerian frequency and $\beta$ is a dimensionless parameter
which is typically assumed to be between 1 and 20. Provided a disc is close to
Toomre instability (i.e. $Q \approx 1$), a disc may only be able to fragment if
the cooling is sufficiently fast ($\beta$ on the order of a few). The critical
value at which gravitational fragmentation occurs, $\beta_{\textup{crit}}$, is
still debated. \cite{Meru:2011a} suggest that the limit may be as high as
$\beta_{\textup{crit}} \approx 30$. More recent studies by \cite{Baehr:2017a}
suggest a value of $\beta_{\textup{crit}} = 3$. 

In this section, we compare the $\beta$-cooling approximation with the cooling
rates which we obtain from Equation \ref{eqn:real_cooling_rate} (which is what
we refer to as {\it actual} cooling). We calculate an effective beta,
$\beta_{\textup{eff}}$, in order to determine whether the assumption of a
constant $\beta$ is a reasonable approximation. Therefore, we define
$\beta_{\textup{eff}}$ as
\begin{equation}
  \beta_{\textup{eff}} = \frac{u}{\dot{u}} \Omega.
\label{eqn:beta_eff}
\end{equation}
where
\begin{equation}
  \dot{u} = \frac{4 \sigma_{\textsc{sb}} T^{4}}
  {\Sigma\left(\tau_{\textsc{r}} + \tau_{\textsc{p}}^{-1}\right)}.
\label{eqn:beta_dudt}
\end{equation}
We emphasise that when calculating $u$ we use the detailed equation of state
used by \cite{Stamatellos:2007b} (see summary in
Section~\ref{sec:efficient_radiative_transfer_methods}).

\begin{figure*}
  \begin{center}
    \includegraphics[width = 1.0\textwidth, trim = 0cm 0cm 0cm 0cm,
    clip=true]{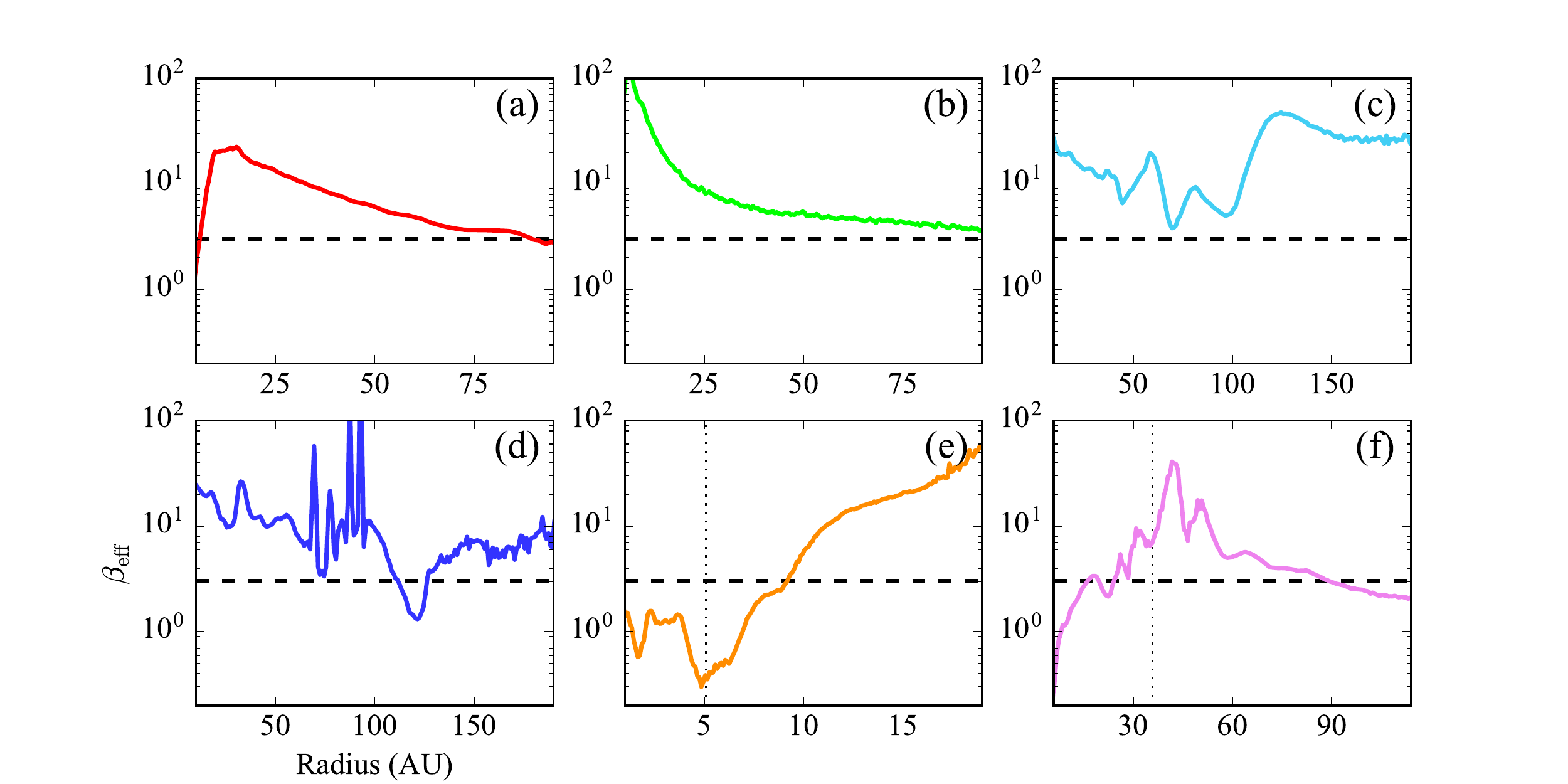}
    \caption
    { 
      Azimuthally-averaged effective $\beta$ at the disc midplane for the
      following snapshots: (a) a low-mass relaxed disc; (b) a high-mass disc;
      (c) a high-mass disc with spirals arms; (d) a high-mass disc with dense
      clumps; (e) a disc with an embedded $1.4 \mjup$ planet; (f) a disc with an
      embedded higher-mass $11 \mjup$ planet. Horizontal dashed lines represent
      $\beta_{\text{eff}} = 3$. Vertical dotted lines represent the radii of
      planets (in the last two cases).
    }
    \label{fig:beta_radial}
  \end{center}
\end{figure*}

\begin{figure*}
  \begin{center}
    \includegraphics[width = 1.0\textwidth, trim = 0cm 0cm 0cm 0cm,
    clip=true]{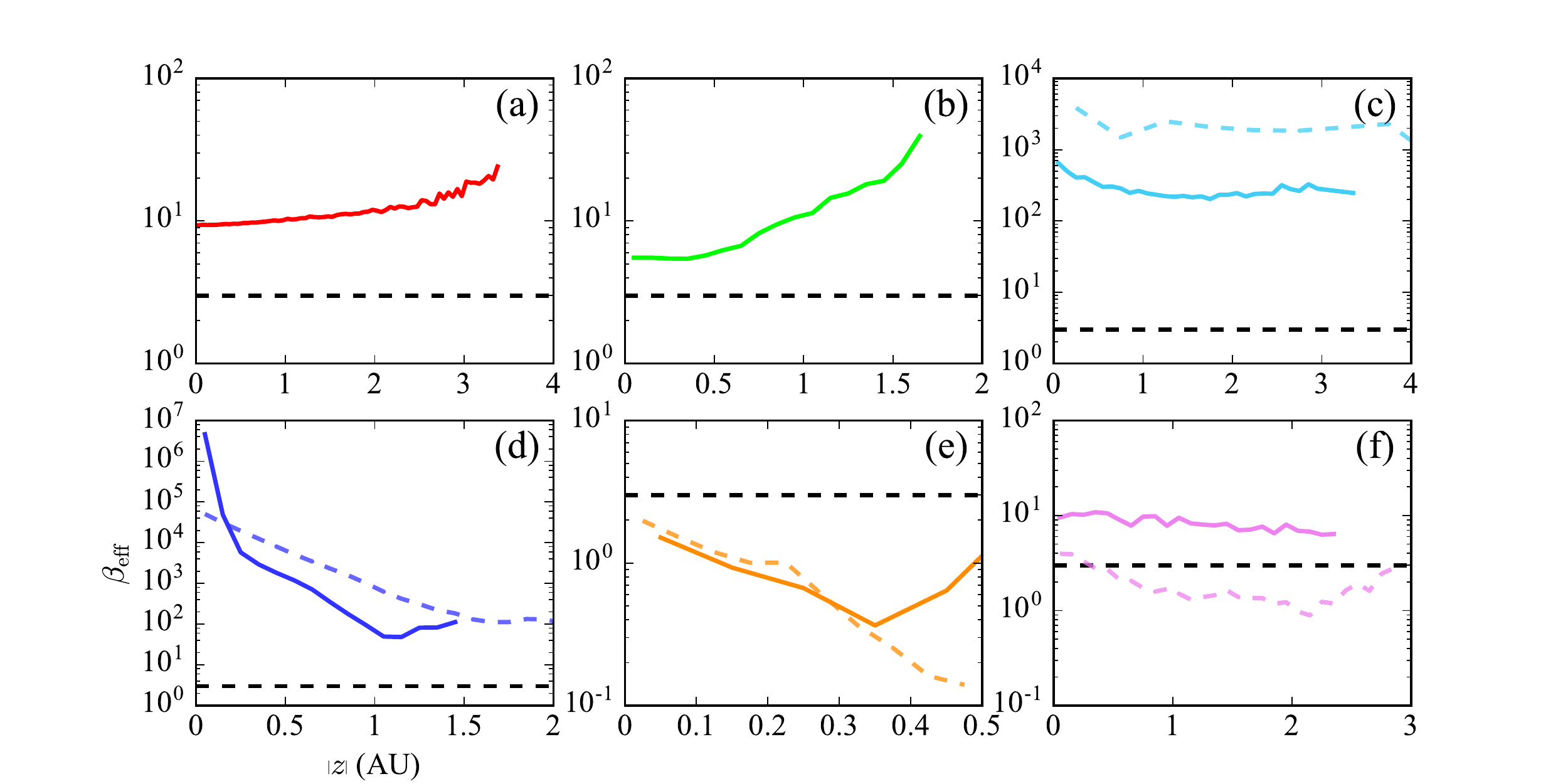}
    \caption
    {
      Effective $\beta$ from the disc midplane to the disc surface for the
      following snapshots: (a) a low-mass relaxed disc (radial annulus $34 < R <
      36$~AU); (b) a high-mass relaxed disc (radial annulus $34 < R < 36$~AU);
      (c) a disc with spirals arms (vertical cylinders with a base with radius
      of 5 AU regions centred within a spiral arm, solid line, and outside
      spiral arms, dashed line); (d) a disc with dense clumps (vertical
      cylinders with a base with radius of 5 AU centred within the densest
      clump, solid line, and the least dense clump, dashed line); (e) a disc
      with an embedded $1.4 \mjup$ planet (radial annuli $4 < R < 6$~AU, solid
      line) and $3 < R < 4$~AU ,dashed line); (f) a disc with an embedded
      higher-mass $11 \mjup$ planet (radial annuli $33 < R < 37$~AU, solid line)
      and $23 < R < 27$~AU, dashed line). Horizontal dashed lines represent
      $\beta_{\text{eff}} = 3$.
    }
    \label{fig:beta_vertical}
  \end{center}
\end{figure*}

\begin{figure*}
  \begin{center}
    \includegraphics[width = 0.99\textwidth, trim = 0cm 6cm 0cm 9cm,
    clip=true]{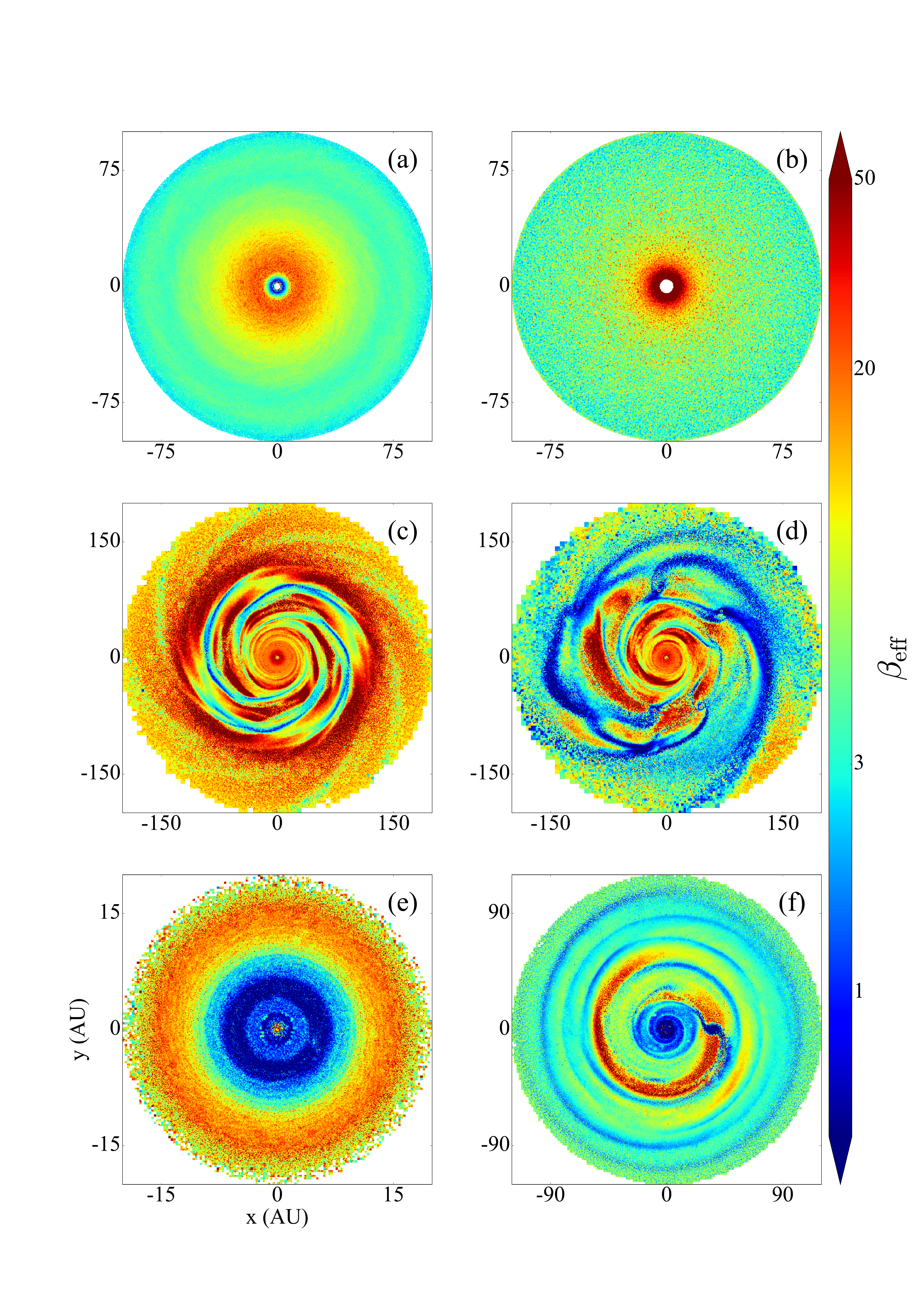}
    \caption
    {
      Effective $\beta$ values at the disc midplane for the following snapshots:
      (a) a low-mass relaxed disc; (b) a high-mass disc; (c) a disc with spirals
      arms; (d) a disc with dense clumps; (e) a disc with an embedded $1.4
      \mjup$ planet; (f) a disc with an embedded higher-mass $11 \mjup$ planet.
      Regions where $\beta_{\textup{eff}}$ is lower cool more efficiently.
      Gravitational instability is typically considered to occur for $\beta < 3$
      provided that the Toomre parameter is also on the order of unity. We show
      that $\beta$ varies across the disc, especially within spiral features and
      dense clumps. As such, it may not be appropriate to assume that $\beta$ is
      constant throughout the disc.
    }
    \label{fig:beta_map}
  \end{center}
\end{figure*}

We present the $\beta_{\textup{eff}}$ that we calculate for the snapshots of
protostellar discs presented in Sections \ref{sec:protostellar_discs} and
\ref{sec:protostellar_discs_with_embedded_planets}. Figure \ref{fig:beta_radial}
shows the azimuthally-averaged $\beta_{\textup{eff}}$ at the disc midplane;
Figure \ref{fig:beta_vertical} shows the value of $\beta_{\textup{eff}}$
vertically towards the surface of the disc at the given regions; Figure
\ref{fig:beta_map} shows colour maps of $\beta_{\textup{eff}}$ at the disc
midplane. We can see that $\beta_{\textup{eff}}$ varies significantly throughout
different regions of each disc, between $\sim 0.1$ and $\sim 200$. 

For the smooth axis-symmetric disc cases that we examine here (Figures
\ref{fig:beta_map}a, b), $\beta_{\textup{eff}}$ is high in the inner disc
regions ($\beta_{\textup{eff}} > 20$) but drops down to $\sim 3$ in the outer
regions. For the disc with the spiral arms (Figure \ref{fig:beta_map}c), the
spirals are regions where $\beta_{\textup{eff}} \sim 1$, hence cooling is
efficient. Thus, spiral arms may be prone to gravitational collapse as thermal
energy generated by the contraction of a forming gas clump can efficiently
escape. The dense, bound clumps in Figure \ref{fig:beta_map}d cool inefficiently
($\beta_{\textup{eff}} \sim 200$), due to being extremely optically thick.

Figure~\ref{fig:beta_map}e shows $\beta_{\textup{eff}}$ for a
disc with a $1.4 \mjup$ embedded planet. $\beta_{\textup{eff}}$ is high in the
outer regions but is low within the planet gap. This may be attributed to the
associated high and low optical depths, respectively, of these regions. For a
disc with an embedded higher-mass $11 \mjup$ planet (Figure
\ref{fig:beta_map}f), the planet induces a high-density spiral wake which cools
fast ($\beta_{\textup{eff}}\sim1$), whereas the gap region cools slowly
($\beta_{\textup{eff}}> 50$). The region around the planet has a low
$\beta_{\textup{eff}}$ ($< 1$) and thus cools more efficiently. 

We see that, as expected, that a region of the disc cools inefficiently (slowly)
when it is optically thin (low-density regions of the disc, e.g in gaps),
efficiently (quickly) when it is just optically thick ($\tau\sim 1$, e.g. in
spirals induced by gravitational instabilities or planets), and again
inefficiently (slowly) when it becomes extremely optically thick (in
clumps/fragments).

We conclude that the actual cooling rate in a protostellar disc varies radially,
vertically and with time as the disc evolves. Significant variations are
observed within dense clumps which form through gravitational fragmentation.
This makes the $\beta$-cooling method a rather crude approximation of the disc
thermal physics when considering highly dynamical systems.

% FIGURE : RADIAL EFFECTIVE BETA------------------------------------------------

% FIGURE : VERTICAL EFFECTIVE BETA----------------------------------------------

% FIGURE : EFFECTIVE BETA HEAT MAPS --------------------------------------------

% ==============================================================================
% DYNAMICAL EVOLUTION COMPARISON
% ==============================================================================
\section{Dynamical evolution comparison}
\label{sec:dynamical_evolution_comparison}

We perform three simulations to demonstrate the differences that the
$\beta$-cooling approximation, the \cite{Stamatellos:2007b}, and the
\cite{Lombardi:2015a} radiative transfer methods exhibit. We simulate a $0.8
\msun$ protostar which is attended by a $0.2 \msun$ disc with surface density
and temperature profiles $\Sigma(R) \propto R^{-1}$ and $T(R) \propto R^{-1/2}$,
respectively. $N \approx 2 \times 10^{6}$ particles represent the disc, which is
heated by a 10 K external radiative field. No heating from the central star is
included. We test the $\beta$-cooling approximation with a value of $\beta = 3$,
a limit at which cooling is efficient enough for gravitational instability to
occur \citep{Rice:2003e}.

\begin{figure*}
  \begin{center}
    \includegraphics[width = 0.99\textwidth, trim = 0cm 22cm 0cm 0cm,
    clip=true]{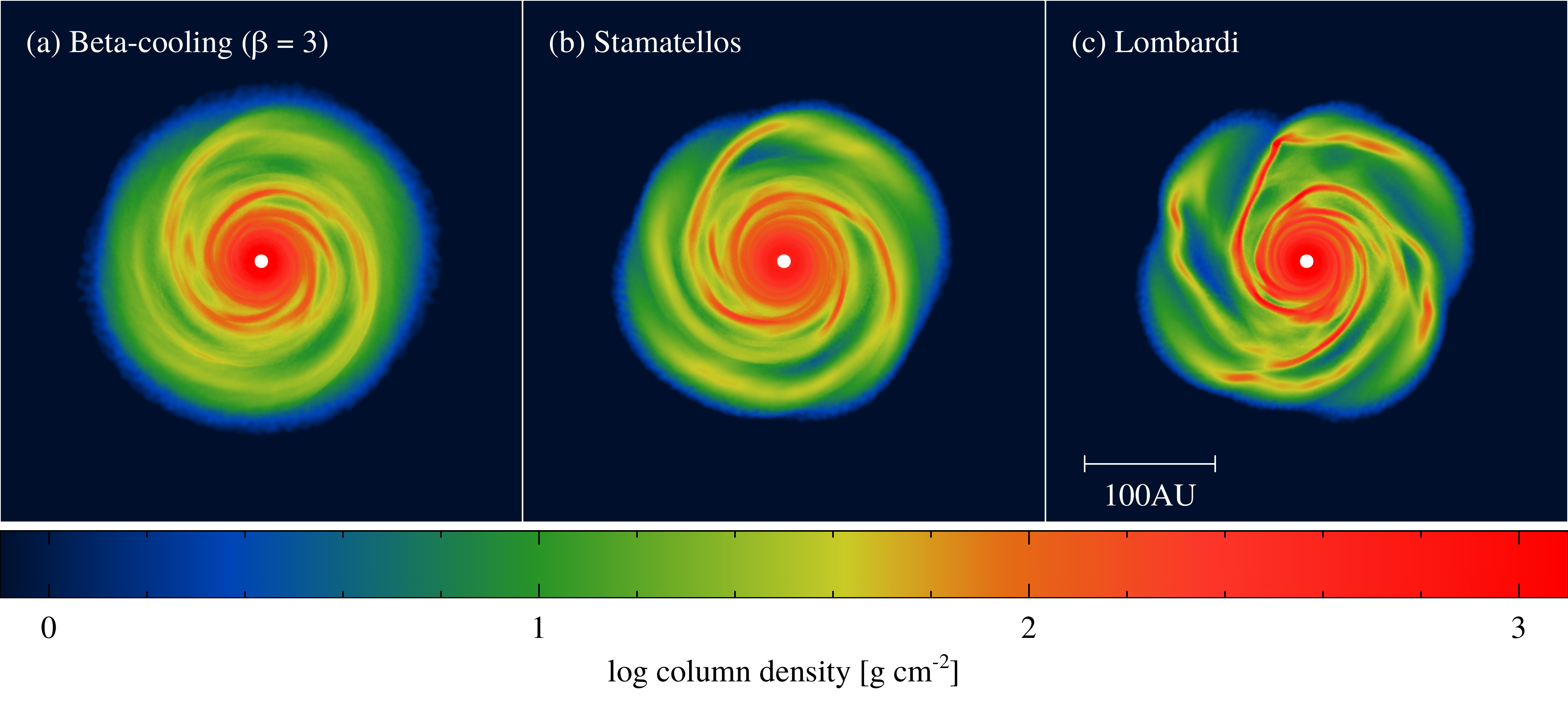}
    \caption
    {
      Surface density plots of a $0.2 \msun$ disc around a $0.8 \msun$ protostar
      after 1.5 kyr of evolution. Panel (a): a disc evolved using the
      $\beta$-cooling approximation with $\beta = 3$. Panel (b): a disc evolved
      using the \protect \cite{Stamatellos:2007b} radiative transfer method.
      Panel (c): a disc evolved using the \protect \cite{Lombardi:2015a} method.
      Each disc becomes gravitationally unstable, but it is clear that the
      Lombardi disc (panel c) is more unstable, demonstrated by the stronger,
      more detailed spiral arms.
    }
    \label{fig:column_density_comparison}
  \end{center}
\end{figure*}

Figure \ref{fig:column_density_comparison} shows the three discs after 1.5 kyr
of evolution using: (a) the $\beta$-cooling approximation; (b) the Stamatellos
radiative transfer method; and (c), the Lombardi radiative transfer method. We
note that whilst all three discs become gravitationally unstable, the
$\beta$-cooling approximation yields a more stable disc than the two radiative
transfer methods. Due to a generally higher estimation of the cooling rate, the
Lombardi method allows the disc to cool more efficiently and develop stronger
spiral arms.

% FIGURE : COLUMN DENSITY PLOTS OF DYNAMICAL EVOLUTION -------------------------

% ==============================================================================
% DISCUSSION
% ==============================================================================
\section{Discussion}
\label{sec:discussion}

We have compared two approximate (but computationally inexpensive) methods to
include radiative transfer in hydrodynamic simulations. These methods use two
different metrics to calculate the optical depth through which the gas heats and
cools: (i) the \cite{Stamatellos:2007b} method uses the gravitational potential
and the density, and (ii) the \cite{Lombardi:2015a} method instead uses the
pressure scale-height.

\begin{figure*}
  \begin{centering}
  \subfloat{\includegraphics[width =0.5\textwidth, trim = 0cm 0cm 0cm 0cm,
  clip=true]{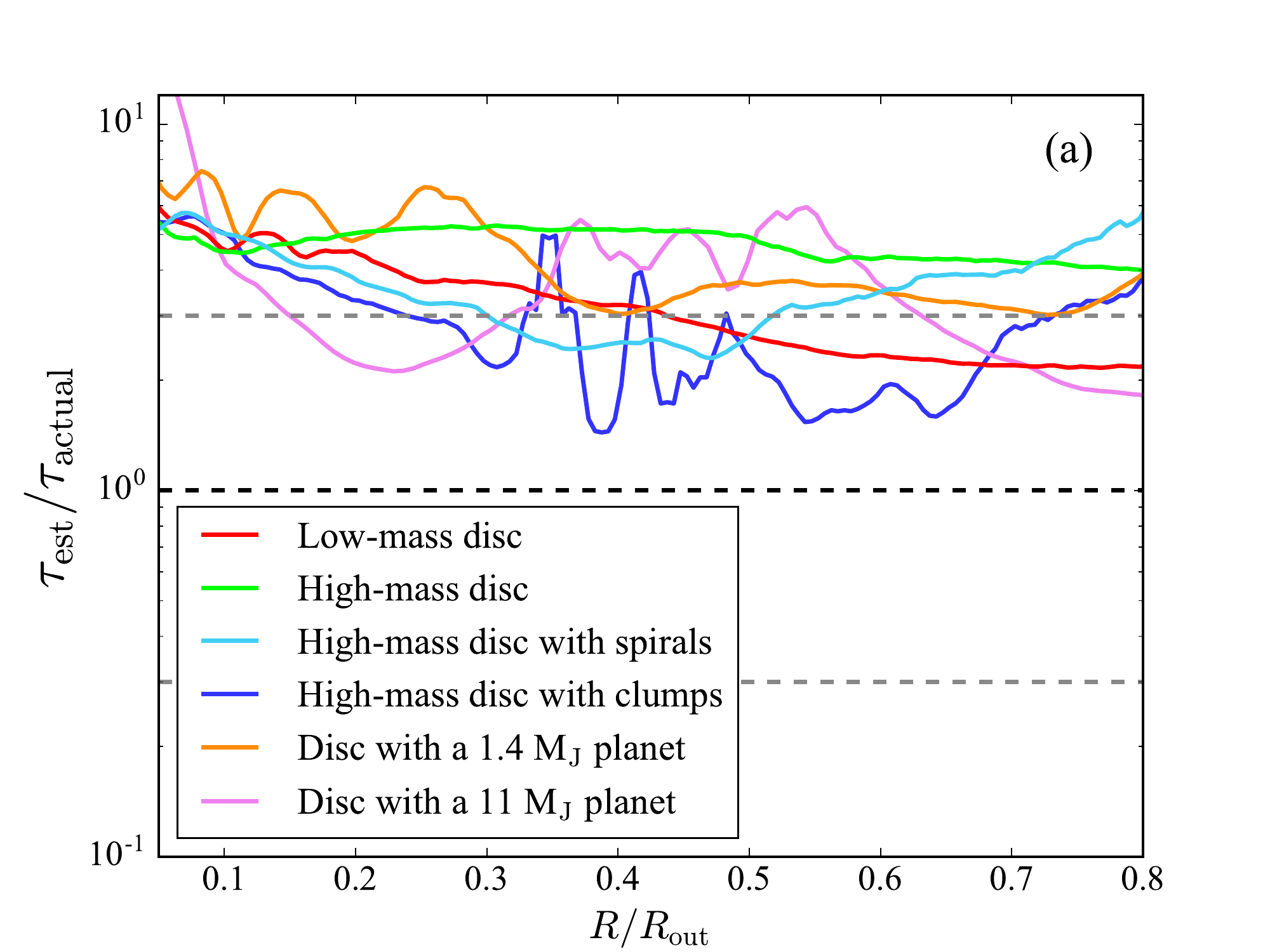}}
  \subfloat{\includegraphics[width =0.5\textwidth, trim = 0cm 0cm 0cm 0cm,
  clip=true]{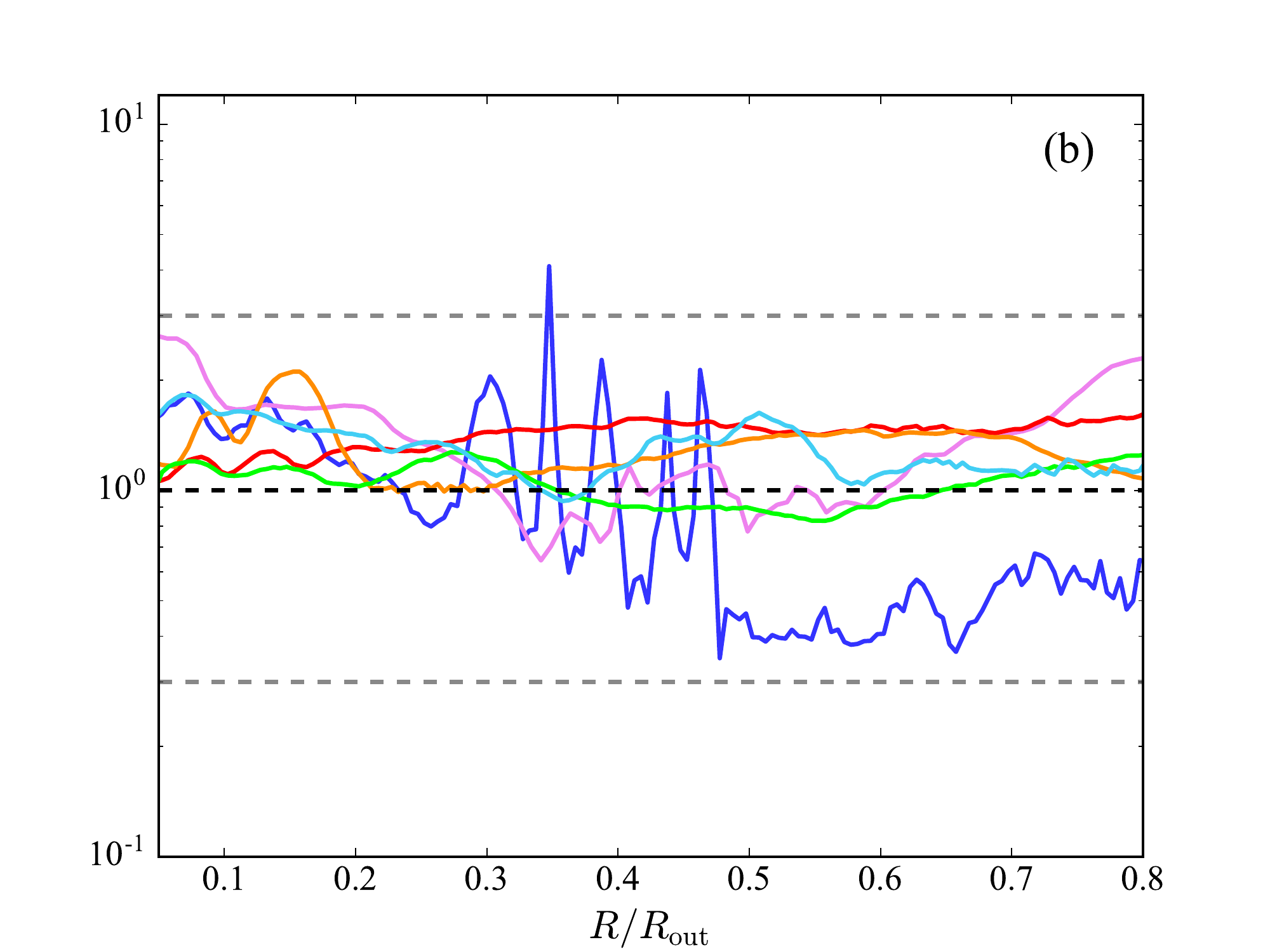}}
  \caption
  {
    The ratio between estimated and actual optical depth for: (a) the \protect
    \cite{Stamatellos:2007b} method; (b) the \protect \cite{Lombardi:2015a}
    method. Various disc configurations are shown. Radii have been normalised to
    the outer radius of each disc. The black dashed lines represent equal values
    of estimated and actual optical depth. The upper and lower grey dashed lines
    represent factors of 3 over- and underestimation respectively. The \protect
    \cite{Lombardi:2015a} metric of estimating optical depths provides better
    accuracy in all cases presented. The optical depth is accurate by a factor
    of less than 3. The \protect \cite{Stamatellos:2007b} method is accurate
    within dense clumps/fragments.
  }
  \label{fig:tau_difference_comparison}
  \end{centering}
\end{figure*}

We find that although both methods yield accurate estimates in the case of
collapsing clouds, the use of the pressure scale-height metric to estimate
optical depths \citep{Lombardi:2015a} is more accurate when considering
protostellar discs. We summarise our results in Figure
\ref{fig:tau_difference_comparison}, which illustrates the difference of optical
depth estimation for the cases we examined in this paper for both methods. Using
the pressure scale-height as a metric, a more accurate estimate of optical depth
(by a factor of 2 or better) and cooling rate is obtained for protostellar discs
in a variety of configurations: low-mass and high-mass discs, with or without an
embedded planet, as well as gravitationally unstable discs which develop spiral
arms and form bound clumps. The \cite{Stamatellos:2007b} method may overestimate
the optical depth by a factor of 10 in some cases, but the \cite{Lombardi:2015a}
method is generally accurate within a factor of 3. Consequently, the
\cite{Stamatellos:2007b} method underestimates the cooling rate in optically
thick protostellar discs, whereas the \cite{Lombardi:2015a} method provides
better accuracy (although generally it also underestimates the cooling rate).
Both methods give accurate estimates in the optically thin regime. 

We also compare the cooling rates in hydrodynamic simulations of discs with
those of the commonly used $\beta$-cooling approximation
\citep[e.g.][]{Gammie:2001a, Rice:2003a}. We find that using a constant value of
$\beta$ for a disc may not be a suitable approximation as this parameter may
vary radially and vertically throughout the disc (between $\sim 0.1$ and $\sim
200$ in the cases that we examined here). It also varies with time as the disc
evolves (e.g. when spiral arms and/or gaps form in the disc), but most
significantly within dense clumps. The approximate radiative transfer methods
discussed previously may be more appropriate to use as, at comparable
computational cost, they are adaptive to the changes that happen as the disc
evolves (e.g. the formation of spiral arms and clumps). Nevertheless, the
$\beta$-cooling approximation is a useful parameterisation that facilitates
greater control in numerical experiments considering the thermal behaviour of a
disc.
 
Many hydrodynamic simulations of protostellar discs (in the context of e.g. disc
evolution, disc fragmentation, disc-planet interactions, planet migration) have
used such approximations for the radiative transfer to avoid excessive
computational cost \citep[e.g][]{Rice:2003e, Lodato:2004a, Clarke:2007a,
Lodato:2007a, Forgan:2009d, Meru:2010b, Stamatellos:2011b, Ilee:2017a}. Their
results need to be seen in the context of the accuracy of the radiative transfer
method used.
 
Studies of disc fragmentation \citep[e.g][]{ Stamatellos:2009a,
Stamatellos:2011d} that use the \cite{Stamatellos:2007b} method may have
underestimated disc cooling by a factor of a few, so that their discs are less
prone to fragmentation. This would mean that even discs with lower masses than
the ones studied by \cite{Stamatellos:2011d} may be able to fragment (i.e. a
disc with mass less than 0.25 M$_{\sun}$ around a 0.7 M$_{\sun}$ star). However,
we should note that the uncertainties in the disc opacities could also be up to
an order of magnitude, i.e. the uncertainty introduced is similar to that of the
\cite{Stamatellos:2007b} method.

Disc simulations using the $\beta$-cooling approximation also suffer from
uncertainties in calculating cooling rates. For discs that start off optically
thin, the cooling is inefficient (i.e. $\beta_{\rm eff}$ is large). The
$\beta_{\rm eff}$ decreases (i.e. the cooling becomes more efficient) as the
density increases in spiral arms and in the region around a planet (i.e. its
circumplanetary disc). If the density continues to increase (i.e. if clumps
form) the cooling becomes inefficient due to the high optical depth, and the
$\beta_{\rm eff}$ increases. The use of a constant $\beta$ misses this variation
of cooling efficiency (both in space and time). Therefore the physics of disc
fragmentation may not be captured appropriately. We demonstrate that a disc
evolved using the $\beta$-cooling approximation, with a value of $\beta = 3$,
results in a more stable disc as compared to similar simulations which employ
the \cite{Stamatellos:2007b} and \cite{Lombardi:2015a} radiative transfer
methods (see Section \ref{sec:dynamical_evolution_comparison}).

In the case of planets embedded in discs, it has been suggested that efficient
cooling promotes gas accretion \citep{Nayakshin:2017b, Stamatellos:2018a} and
dust accretion \citep{Humphries:2018a} onto the planet. Therefore, cooling rates
may affect the mass growth of planets, their metallicity, and their associated
circumplanetary discs. This in turn results in different migration rates, final
masses and orbital characteristics for these planets e.g. as seen in
\cite{Stamatellos:2015a} in comparison with \cite{Baruteau:2011a}
\citep[see][]{Stamatellos:2018a}.
 
% FIGURE : FULL TAU DIFFERENCE COMPARISON --------------------------------------

% ==============================================================================
% CONCLUSION
% ==============================================================================
\section{Conclusion}
\label{sec:conclusion}

Approximate radiative transfer methods are useful due to their computational
efficiency, but they should be treated with caution as radiative transfer may,
in many cases, fundamentally affect the evolution of an astrophysical system.
The \cite{Lombardi:2015a} method (that uses the pressure scale-height to
calculate optical depths) is more accurate than the \cite{Stamatellos:2007b}
method (that uses the gravitational potential and the gas density as a proxy for
optical depths) for disc simulations. Both methods behave accurately for
spherical geometries (i.e. collapsing clouds or clumps in discs). When used for
modelling protostellar discs, both methods are more accurate than the
$\beta$-cooling approximation (at similar computational cost), which
nevertheless is a good tool for controlled numerical experiments of disc
thermodynamics.

% ==============================================================================
% ACKNOWLEDGEMENTS
% ==============================================================================
\section*{Acknowledgements}
\label{sec:acknowledgements}

The authors would like to thank Richard Booth and David Hubber for the
correspondence on the implementation of the methods used within this paper as
well as useful comments. Thanks are given to the anonymous referee for
constructive suggestions. Surface density plots were produced using the
\textsc{splash} software package \citep{Price:2007b}. AM is supported by STFC
grant ST/N504014/1. DS is partly supported by STFC grant ST/M000877/1. This work
used the DiRAC Complexity system, operated by the University of Leicester IT
Services, which forms part of the STFC DiRAC HPC
Facility\footnote{\texttt{http://www.dirac.ac.uk}}. This equipment is funded by
BIS National E-Infrastructure capital grant ST/K000373/1 and STFC DiRAC
Operations grant ST/K0003259/1. DiRAC is part of the UK National
E-Infrastructure.

% ==============================================================================
% BIBLIOGRAPHY
% ==============================================================================
\bibliography{rtimpd}{}
\bibliographystyle{mnras}

% Don't change these lines
\bsp	% typesetting comment
\label{lastpage}
\end{document}